\documentclass[namedreferences]{solarphysics}

\usepackage[hyperref,optionalrh,natbib]{spr-sola-addons} 
\usepackage{graphicx}        
\usepackage{color}           

\begin{document}
\begin{article}

\begin{opening}

\title{Observational Signatures of Waves and Flows in the Solar Corona}

\author{I.~\surname{De Moortel}$^{1}$\sep
        P.~\surname{Antolin}$^{2}$\sep
        T.~\surname{Van Doorsselaere}$^{3}$      
       }
\runningauthor{De Moortel et al.}
\runningtitle{Observational Signatures of Waves and Flows}

   \institute{$^{1}$ School of Mathematics and Statistics, University of St Andrews, St Andrews, Fife, KY16 9SS, U.K.
                     email: \url{ineke.demoortel@st-andrews.ac.uk}\\ 
              $^{2}$ National Astronomical Observatory of Japan, 2-21-1 Osawa, Mitaka, Tokyo 181-8588, Japan 
                     email: \url{patrick.antolin@nao.ac.jp} \\
                $^{3}$ Centre for mathematical Plasma Astrophysics, Mathematics Department, KU Leuven, Celestijnenlaan 200B bus 2400, 3001 Leuven, Belgium
                     email: \url{tom.vandoorsselaere@wis.kuleuven.be} \\
             }

\begin{abstract}
Propagating perturbations have been observed in extended coronal loop structures for a number of years but the interpretation either in terms of slow (propagating) magneto acoustic waves or as quasi-periodic upflows remains unresolved. This paper uses forward modelling to construct observational signatures associated with a simple slow magneto acoustic wave or periodic flow model. Observational signatures are computed for the 171 \AA\ Fe~{\sc{ix}} and the 193 \AA\ Fe~{\sc{xii}} spectral lines. Although there are many differences between the flow and wave models, we do not find any clear, robust observational characteristics which can be used in isolation ({\it {\it i.e.}}~which do not rely on a comparison between the models). For the waves model, a relatively rapid change of the average line widths as a function of (shallow) line-of-sight angles is found whereas for the flow model, the ratio of the line width amplitudes to the Doppler velocity amplitudes is relatively large. The most robust observational signature found is the fact that the ratio of the mean to the amplitudes of the Doppler velocity is always larger than one for the flow model. This ratio is substantially bigger for flows than for waves and for the flows model used in the study is exactly the same in the 171 \AA\ Fe~{\sc{ix}} and the 193 \AA\ Fe~{\sc{xii}} spectral lines. However, these potential observational signatures need to be treated cautiously as they are likely to be model-dependent.
\end{abstract}
\end{opening}

\section{Introduction}\label{sec:intro}

Since the advent of high resolution imagers, there have been many observations of intensity (density) perturbations travelling along coronal loops.  However, from the very outset, two different interpretations of these `propagating coronal disturbances' (PCDs) could be found in the literature: both a propagating, slow magneto-acoustic wave and periodic upflows can lead to periodic density perturbations, which would be observed as propagating, periodic intensity variations by imaging instruments. For a comprehensive review of these propagating (coronal) disturbances, we refer the reader to, for example, \cite{IDM09}, \cite{review:DeMoortelNakariakov2012} or \cite{Banerjee2011}.

Initial reports by \cite{Schrijver99} and \cite{Winebarger02} found lower propagation speeds (of the order of 40 km s$^{-1}$), and hence supported an interpretation in terms of a (quasi-)periodic flow model. Subsequently however, various authors reported similar disturbances in coronal plumes ({\it e.g.}~\citealt{Ofman97,DeForest1998,Banerjee2000}) and large coronal (fan) loops at the edges of active regions ({\it e.g.}~\citealt{Berghmans1999,IDM00,IDM2002b,IDM2002a}), propagating at speeds close to the local sound speed, leading to the alternative interpretation in terms of slow magneto-acoustic waves. Combined with theoretical modelling which explained the decay of the perturbations in terms of thermal conduction ({\it e.g.}~\citealt{Ofman00,IDM03,IDM04}), the slow wave model became widely accepted. However, more recently, additional spectral observations provided by Hinode/EUV Imaging Spectrometer (EIS, \citealt{Culhane2007})  has reopened the debate on whether to interpret the observed PCDs as a slow propagating wave or as quasi-periodic upflows. Not only are perturbations in intensity observed, but also in other parameters such as Doppler velocity, line-widths and red-blue asymmetries, which appear consistent with an interpretation in terms of quasi-periodic upflows ({\it e.g.}~\citealt{Sakao2007,Doschek2008,DelZanna2008,Hara2008,Harra2008,McIntosh2009a,McIntosh2009b,DePontieu2009,DePontieu2011,DePontieu2010,He2010,Bryans2010,Tian2011,Ugarte-Urra2011,Warren2011}). Again we refer the reader to \cite{review:DeMoortelNakariakov2012} as well as to \cite{McIntosh2012REV} for a more thorough review of the literature.

Following this new series of observations, however, \cite{Verwichte2010} and \cite{Wang2012} pointed out that it would still be possible to interpret the combined imagining and spectral observations in terms of slow magneto-acoustic waves, leaving the interpretation of the observed propagating coronal disturbances (PCDs) inconclusive.  We also refer the interested reader to \cite{Peter2010} for a discussion on the observed asymmetries in EUV emission lines. Both the slow wave and the periodic upflow models can partially explain the observations but neither can currently account for all of the observed properties. This is perhaps most clearly illustrated by \cite{Wang2009} and  \cite{DePontieu2010}, who analyse exactly the same dataset but arrive at a different interpretation, with the first explaining the observed PCDs in terms of slow magneto acoustic waves and the latter using a quasi-periodic upflow model. More recently, a ``dual'' model has been suggested, where quasi-periodic flows at the very footpoints of the loops generate a slow magneto-acoustic wave which travels further along the loops (at the local sound speed) as {\it e.g.}~in the observations of \cite{Nishizuka2011} or the model of \cite{Ofman2012} and \cite{Wang2013}.

Although the debate might seem a mainly semantic one, there are important underlying physical implications. If the observed PCDs are indeed propagating slow magneto acoustic waves, seismology can be used to derive local plasma parameters. Using this interpretation, for example, \cite{Marsh2009a} and \cite{Marsh2009b} inferred the local plasma temperature and \cite{VanDoorsselaere2011} estimated the local thermal conduction coefficient and polytropic index.  If, on the other had, the PCDs are indeed quasi-periodic upflows, they could play a significant role in the coronal mass cycle given their abundant and continuous presence in large coronal loops and fans (see {\it e.g.}~\citealt{McIntosh2009b,McIntosh2010,McIntosh2012,DePontieu2011}). 

As the observational data have been pushed to their (current) limits, we take a different approach in this paper to try and resolve this debate. We use theoretical modelling of both a propagating slow magneto acoustic wave and a periodic flow, combined with forward modelling to study the possible observational signatures of each model (see {\it e.g.}~\citealt{IDM08,Owen2009,Antolin2013}). The model setup and forward modelling process are described in more detail in Section \ref{sec:model}, followed by a description of the initial value Fe~{\sc{ix}} results (Section \ref{sec:results_ini}), the Fe~{\sc{ix}} results for a harmonic driver (Section \ref{sec:results_hmn}) and the results for the Fe~{\sc{xii}} line (Section \ref{sec:results_fe12}). A discussion of the results and conclusions are presented in Sections \ref{sec:disc} and \ref{sec:concl}.

\section{Model Setup and Forward Modelling}\label{sec:model}

\begin{figure}[t]
\centering
\scalebox{.25}{\includegraphics{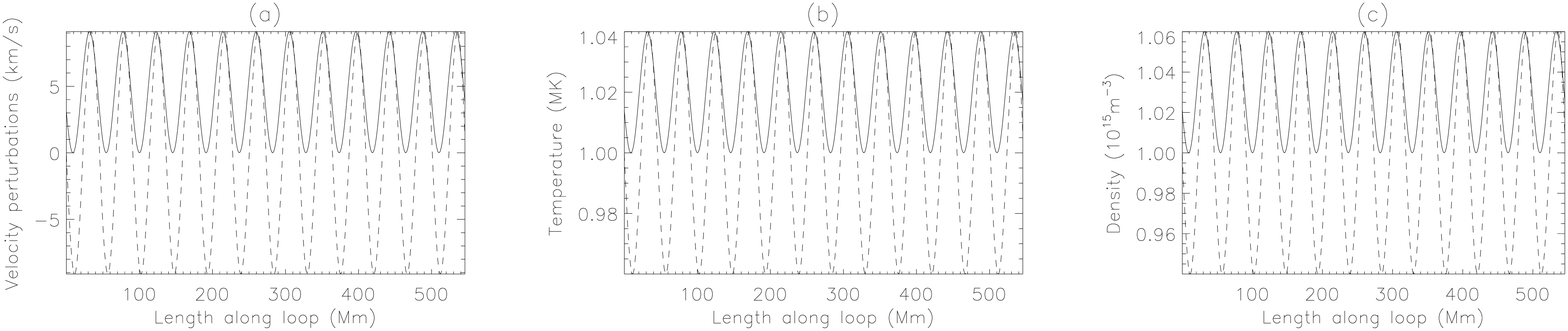}}
\caption{Evolution of the (a) velocity, (b) temperature and (c) density as a function of distance along the loop for the flow (solid lines) and wave (dashed lines) models.}
\label{fig:model}
\end{figure}

In order to determine observationally distinguishable signatures for propagating waves and periodic flows, we construct a simple, 2D numerical model using LareXd \citep{Arber2001}. The uniform, background medium has a typical coronal temperature of $T_0=1$ MK and density of $n_0=10^{15}$ m$^{-3}$. A propagating (longitudinal) wave is modelled by using a lower-boundary driver of the form
$$v(x,y=0,t)=A\sin(\omega_{{\rm wave}} t)\,,$$
whereas a flow is modelled by 
$$v(x,y=0,t)=A\sin^2(\omega_{{\rm flow}} t)\,.$$
Here $A$ is the amplitude and $\omega$ is the frequency. Small amplitudes are used to avoid non-linear interactions. Note that since, 
\begin{equation}
\sin^2 \omega = {1 \over 2}(1-\cos 2 \omega)
\label{eq:superpos}
\end{equation}
we use $\omega_{{\rm flow}}=\omega_{{\rm wave}}/2$, where $\omega_{{\rm wave}}=2\pi$ (corresponding to a period of 300 s in dimensional units), to obtain perturbations with the same frequency (see Figure~\ref{fig:model}). However, this immediately tells us that the periodic flow can be decomposed into a (periodic) wave and a steady background flow with the same amplitude. The boundary-driven velocity perturbations travel along the loop at the local sound speed (for both the wave and flow perturbations), generating associated temperature and density perturbations. As can be seen from Figure~\ref{fig:model}, the perturbations associated with the wave mode (dashed lines) oscillate around the equilibrium values whereas the flow perturbations are always greater than the background values. In this sense, there is already not a lot of physical difference between the wave or flow simulations, since the periodic flow simulation is just a propagating sound wave on top of a background flow. From our simulations, we have not found evidence for the existence of periodic flows that propagate at a speed less than the sound speed, and how they could be initiated. The periodic inflow at the footpoint automatically triggers an upwardly propagating sound wave.

\begin{figure}[t]
\centerline{\hspace*{0.015\textwidth}
               \includegraphics[width=0.2\textwidth,clip=]{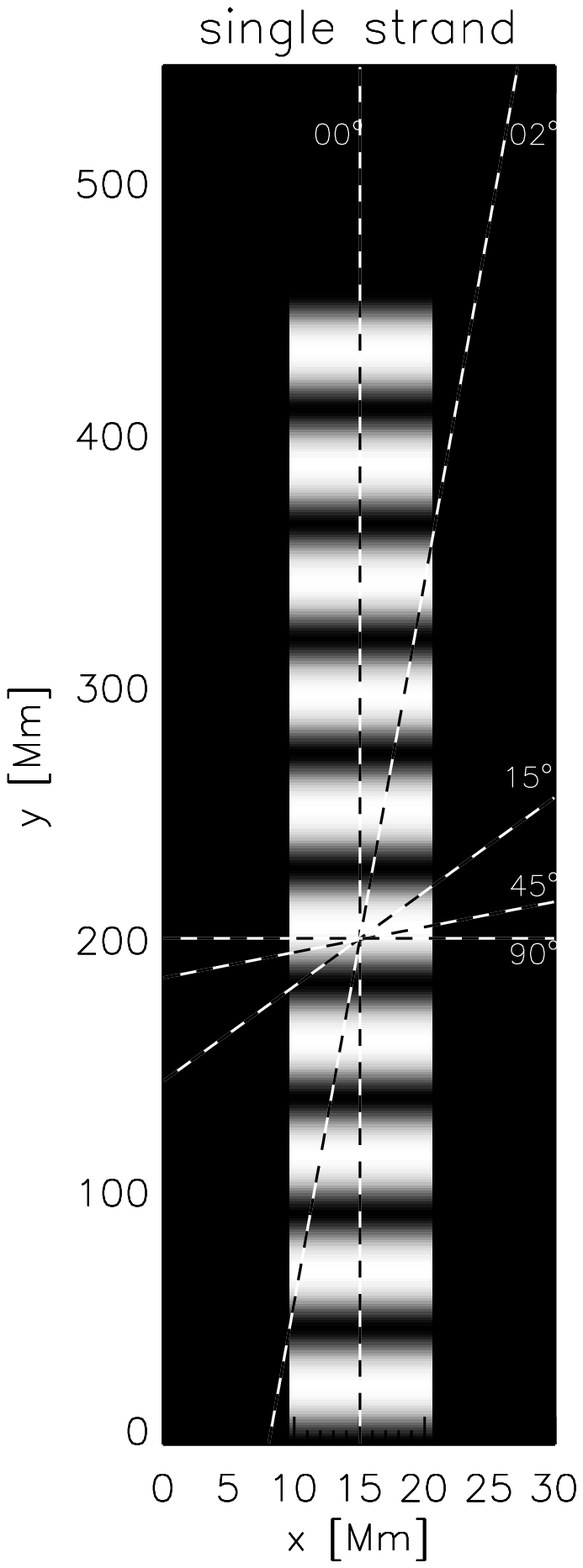}
               \hspace*{-0.03\textwidth}
               \includegraphics[width=0.2\textwidth,clip=]{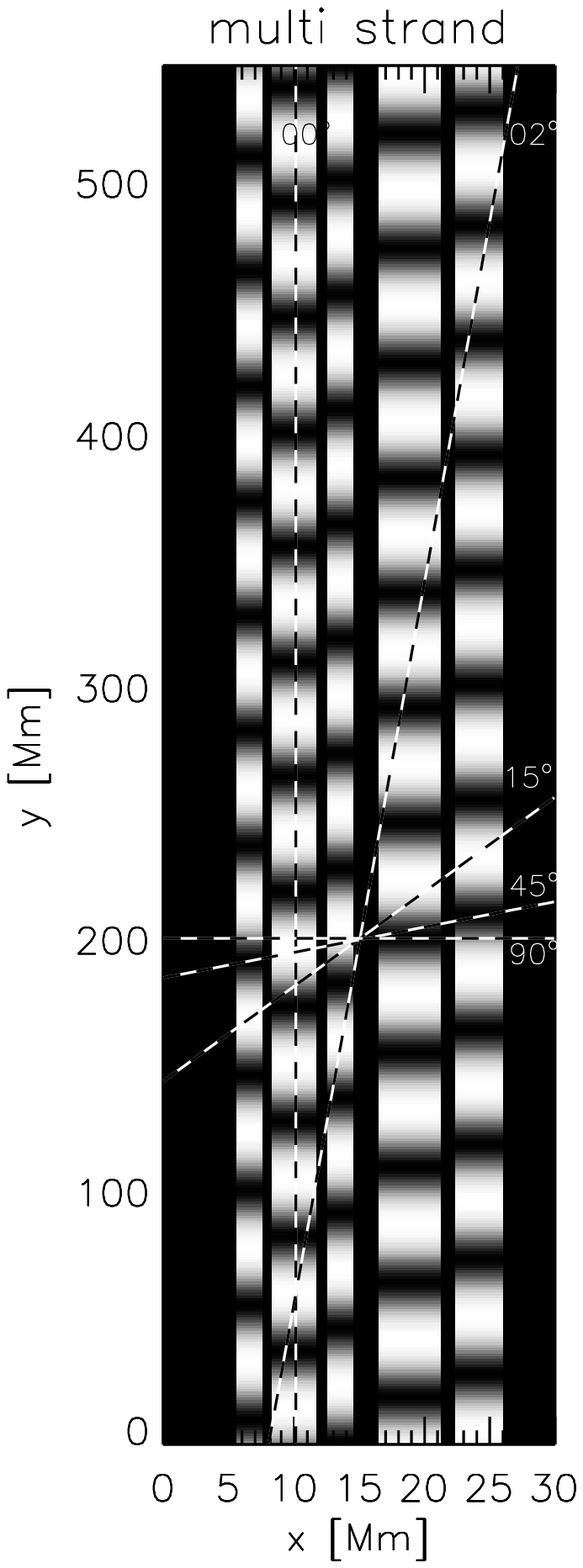}
              }
     \vspace{-0.55\textwidth}  
     \centerline{\bf \small    
      \hspace{0.4 \textwidth}  \color{black}{(a)}
      \hspace{0.13\textwidth}  \color{black}{(b)}
         \hfill}
 \vspace{0.5\textwidth}   
 \caption{Snapshots showing (a) the initial value single-strand (at $t=3000$ s) and (b) the steady-state multistrand perturbations and the rays used to trace at different angles (dashed lines).}
\label{fig:rays}
\end{figure}

Two different versions of the model are created. The first is a simple, single strand of perturbations embedded in a uniform background as shown in Figure~\ref{fig:rays}(a). The second model represents a ``multistrand" loop, consisting of five small sections of perturbations all with the same periods but slightly out of phase, as shown in Figure~\ref{fig:rays}(b). 

Observational signatures are subsequently forward modelled, using the FoMo forward modelling code\footnote{https://wiki.esat.kuleuven.be/FoMo} (see \cite{Antolin2013} for a description of the forward modelling code). Note that we are not interested in the absolute values of the observables but in the relative differences between the signatures of the wave and flow models. We model the Fe~{\sc{ix}} 171.073 \AA\ and Fe~{\sc{xii}} 193.509 \AA\ spectral lines, corresponding to the dominant lines in the SDO/AIA 171 {\AA} and 193 {\AA} filters, respectively \citep{Lemen2012}. Finally, the effect of the line-of-sight (LOS) angle is incorporated by integrating along different `rays', as shown in Figure~\ref{fig:rays}  by the dashed lines. A $0^\circ$ LOS angle corresponds to a loop aligned with the LOS ({\it i.e.}~the perturbations are travelling directly towards the observer) whereas a $90^\circ$ LOS angle corresponds to a loop perpendicular to the LOS ({\it i.e.}~there are no velocity perturbations aligned with the LOS). We will denote the LOS angle as $\theta$. For each of the numerical simulations, we present Doppler velocity, line width and the goodness-of-fit measure $\chi^2/\chi_0^2$ as a function of time. Here $\chi^2/\chi_0^2$ is a measure of how closely the Gaussian fit to the spectral line matches the corresponding Gaussian fit for the plasma at rest. Unless otherwise mentioned, a single Gaussian fit to the spectral lines is used.

\section{Initial Value Results}\label{sec:results_ini}

In this section, we present initial value results ({\it i.e.}~there are no perturbations in the domain  at $t=0$)  where the perturbations are driven at the bottom boundary and propagate into the domain. Hence, the system changes from initially being at rest to a system containing a wave or flow. The simulations are stopped before they reach the upper boundary as can be seen in the single-strand example in Figure~\ref{fig:rays}(a).

\subsection{Single Strand Wave and Flow}

\begin{figure}[t]
\centerline{\hspace*{0.015\textwidth}
               \includegraphics[width=0.4\textwidth,clip=]{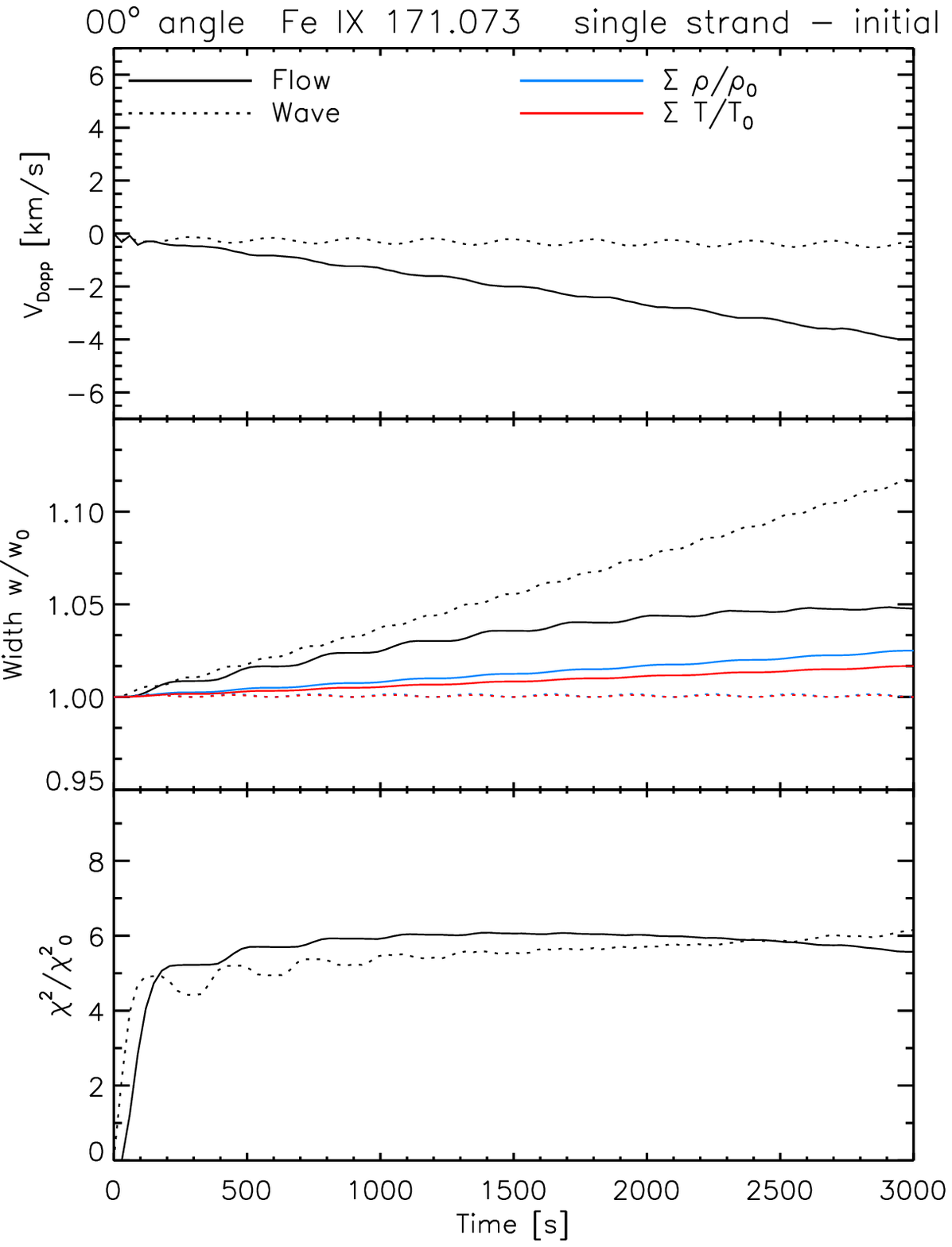}
               \hspace*{-0.03\textwidth}
               \includegraphics[width=0.4\textwidth,clip=]{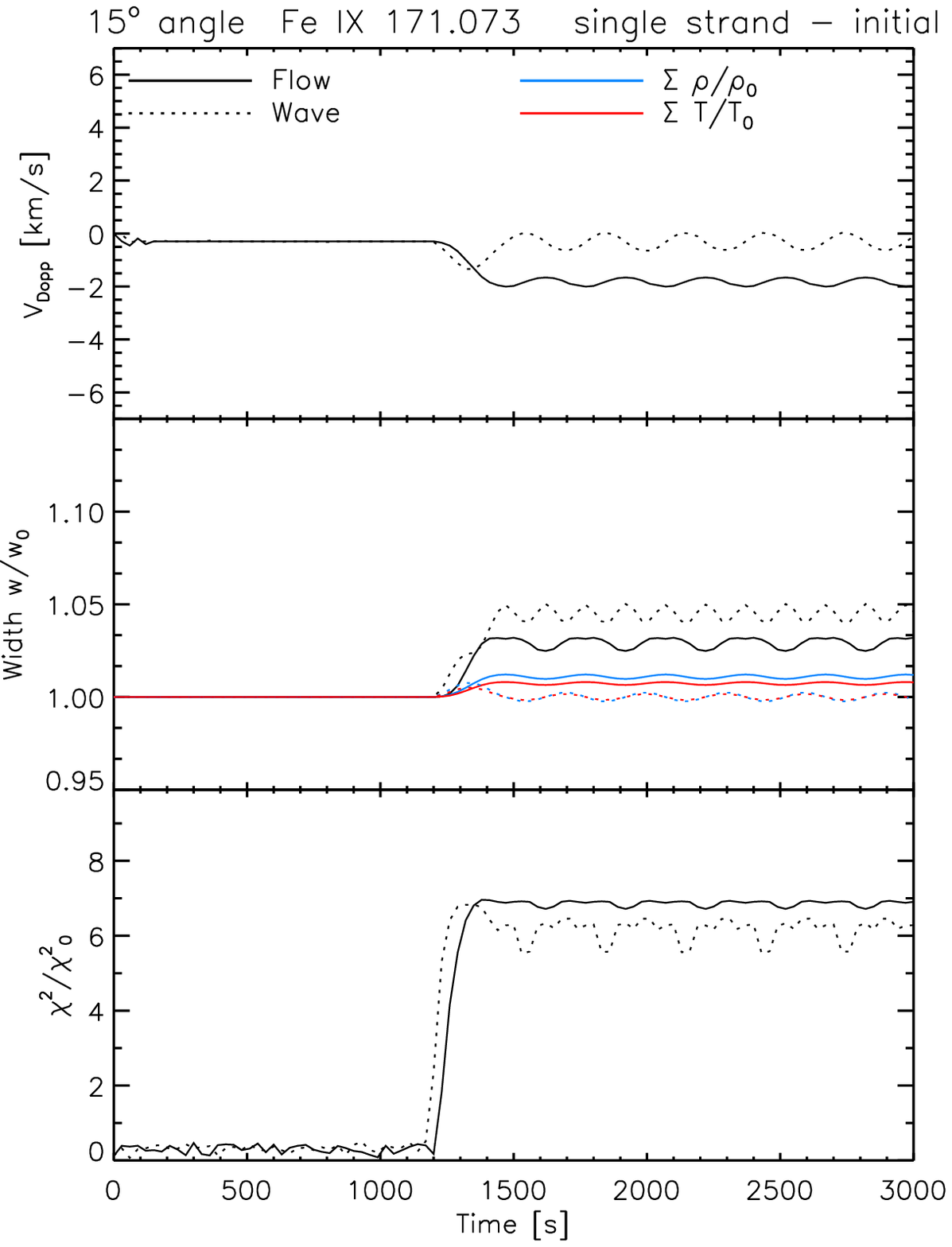}}
 \vspace{-0.57\textwidth}  
     \centerline{\bf \small    
      \hspace{0.31 \textwidth}  \color{black}{(a)}
      \hspace{0.31\textwidth}  \color{black}{(b)}
         \hfill}
  \vspace{0.55\textwidth}    
\centerline{\hspace*{0.015\textwidth}               
                \includegraphics[width=0.4\textwidth,clip=]{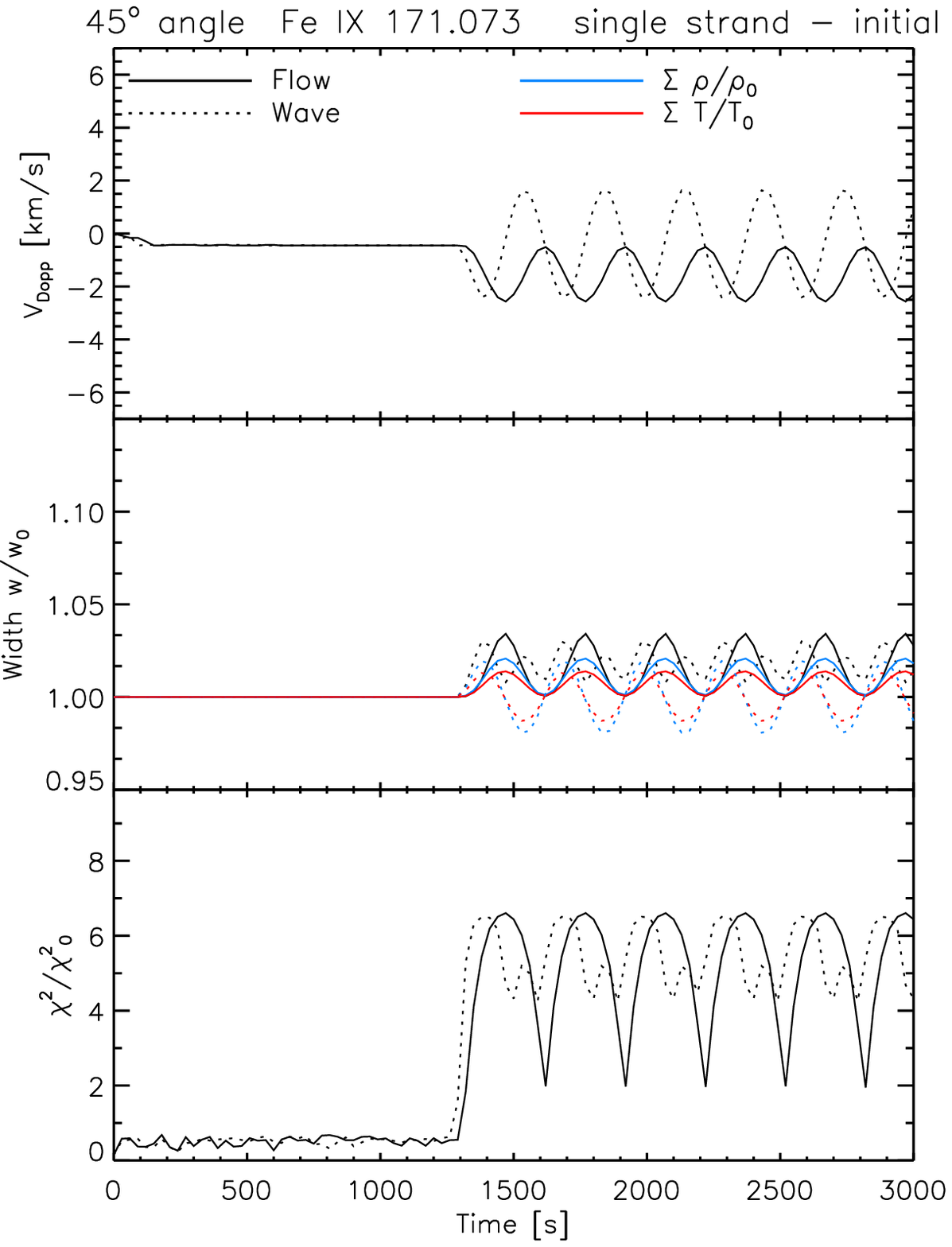}
               \hspace*{-0.03\textwidth}
               \includegraphics[width=0.4\textwidth,clip=]{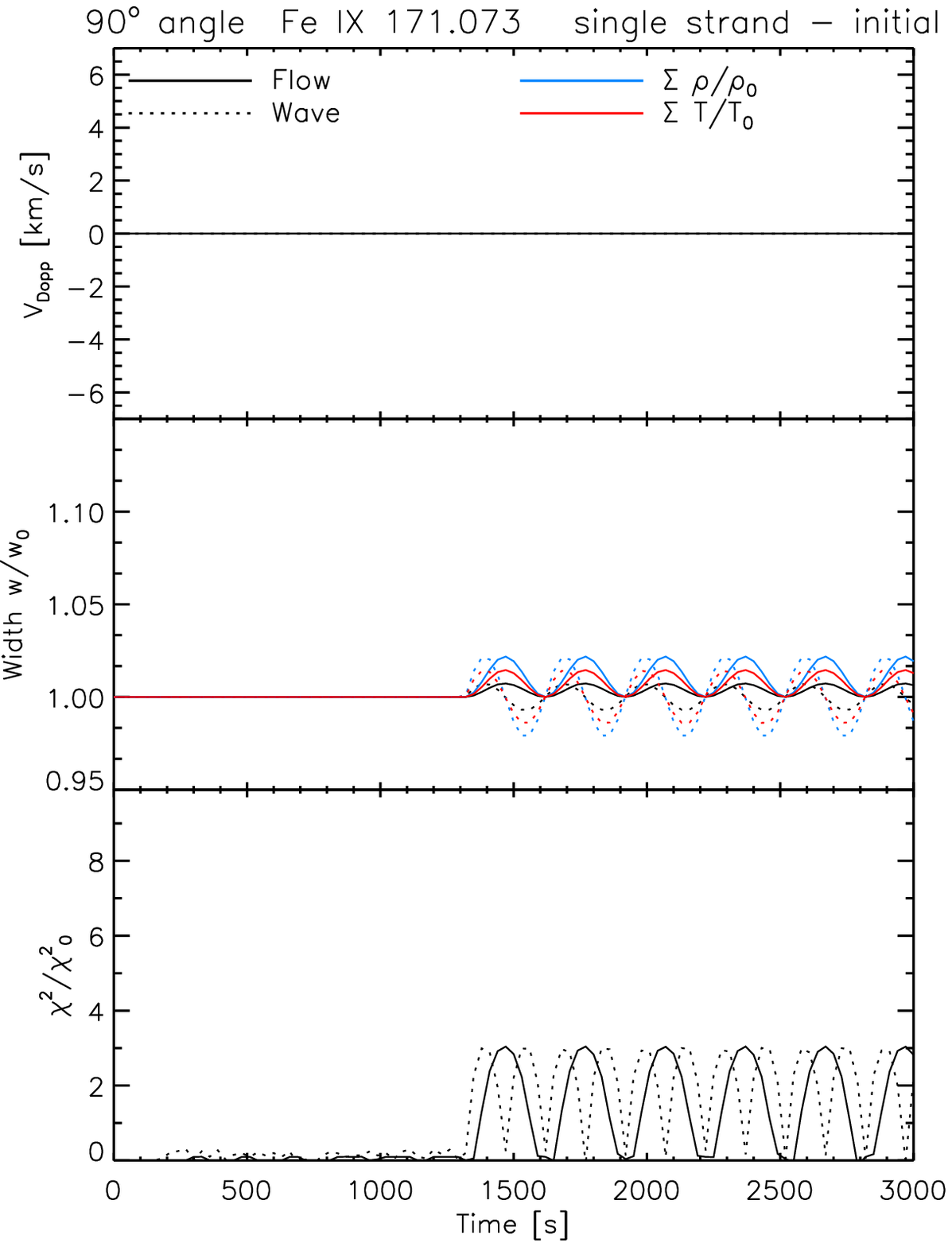}
              }
     \vspace{-0.57\textwidth}  
     \centerline{\bf \small    
      \hspace{0.31 \textwidth}  \color{black}{(c)}
      \hspace{0.31\textwidth}  \color{black}{(d)}
         \hfill}
 \vspace{0.52\textwidth}   
\caption{Time evolution (initial value, single-strand case) of the 171 \AA\ Fe~{\sc{ix}} Doppler velocity (top graphs in four panels), line widths (middle graphs) and  the goodness-of-fit measure $\chi^2/\chi_0^2$ (bottom graphs) for tracing rays at angles of (a) $0^\circ$, (b) $15^\circ$, (c) $45^\circ$, and (d) $90^\circ$. The relative density (blue) and temperature (red) are overplotted for comparison. The results for the flow model are represented by the solid lines and the waves model by the dashed lines.}
\label{fig:profiles171_inival_single}
\end{figure}


\begin{figure}[t]
\centerline{\hspace*{0.015\textwidth}
               \includegraphics[width=0.4\textwidth,clip=]{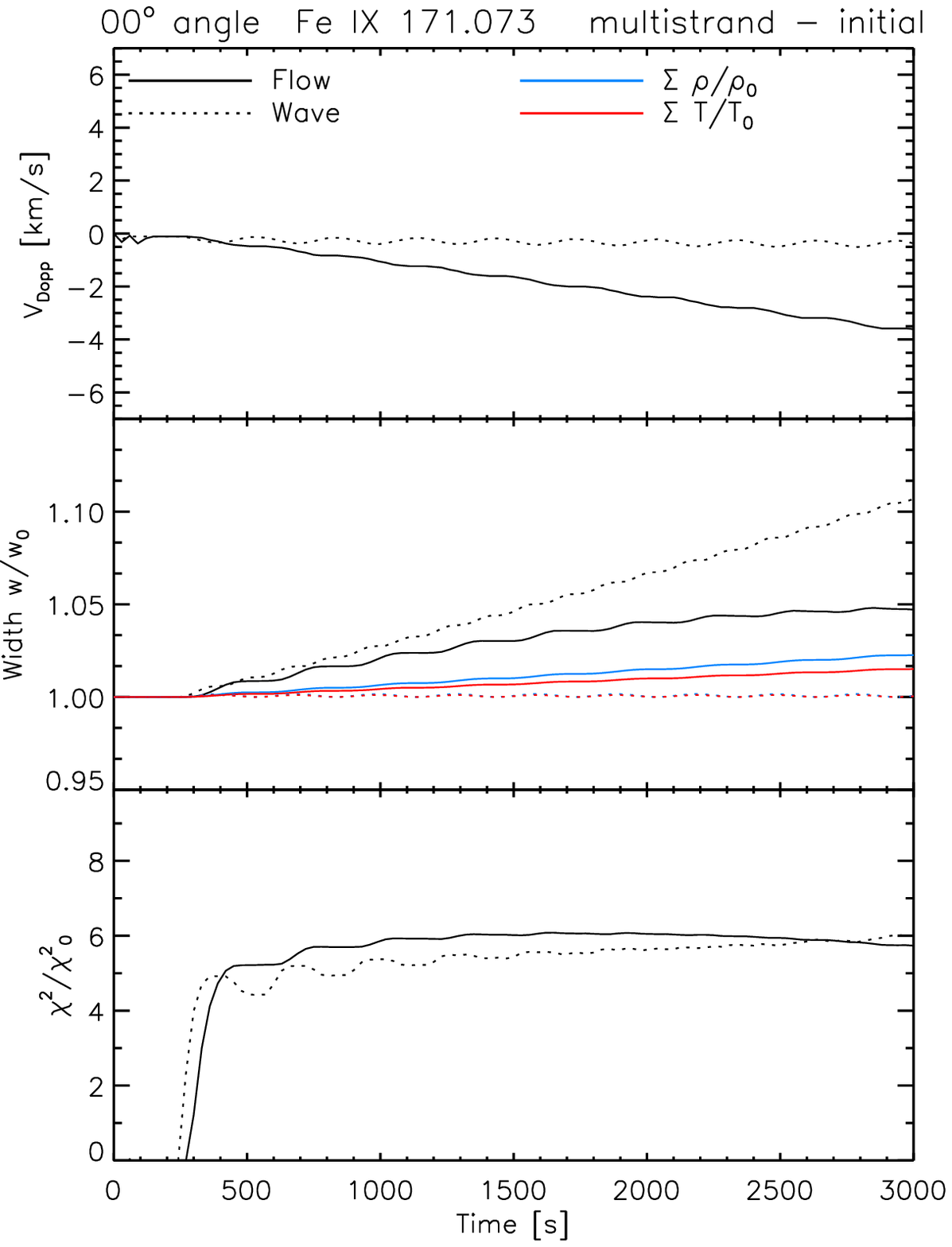}
               \hspace*{-0.03\textwidth}
               \includegraphics[width=0.4\textwidth,clip=]{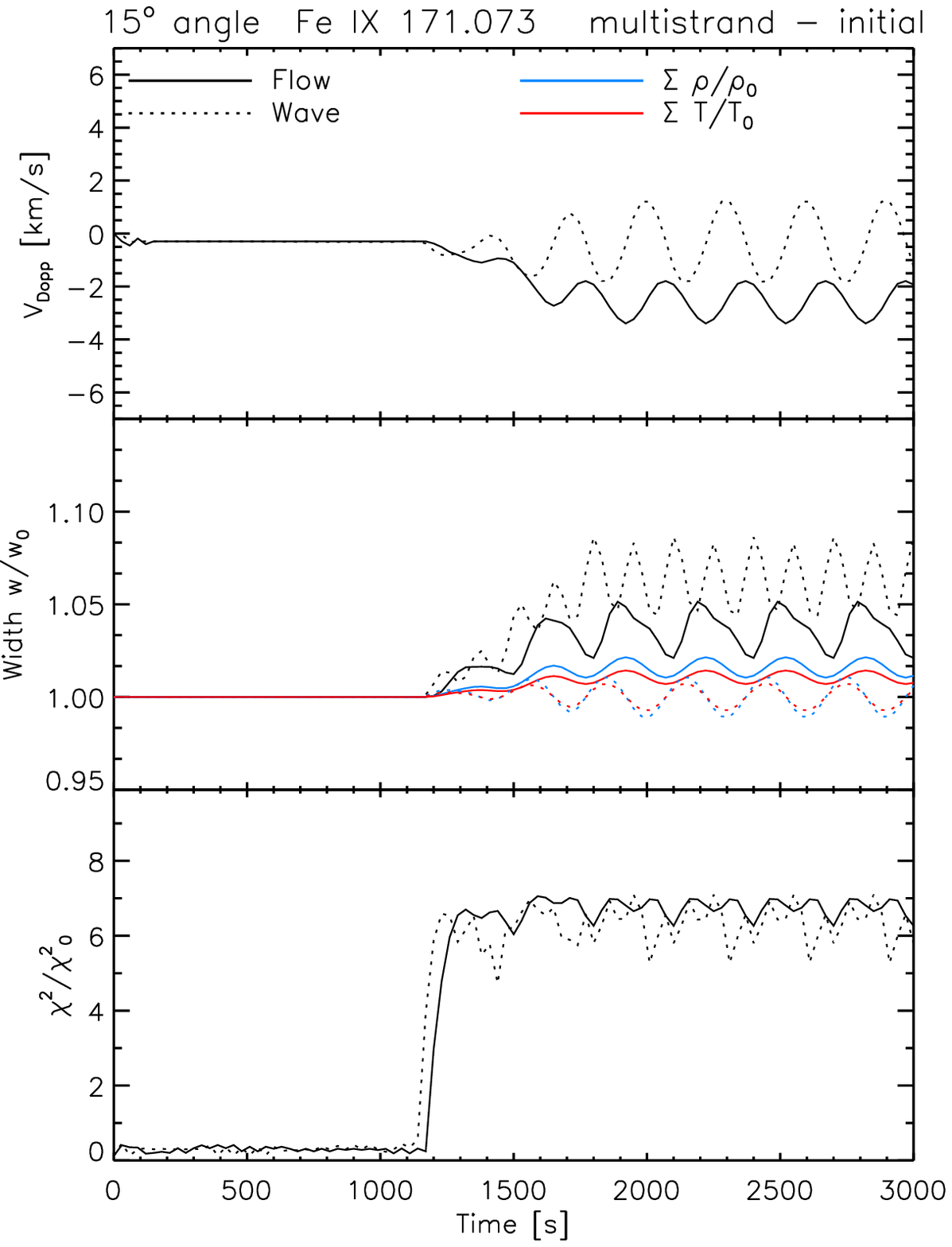}}
 \vspace{-0.57\textwidth}  
     \centerline{\bf \small    
      \hspace{0.31 \textwidth}  \color{black}{(a)}
      \hspace{0.31\textwidth}  \color{black}{(b)}
         \hfill}
  \vspace{0.55\textwidth}    
\centerline{\hspace*{0.015\textwidth}               
                \includegraphics[width=0.4\textwidth,clip=]{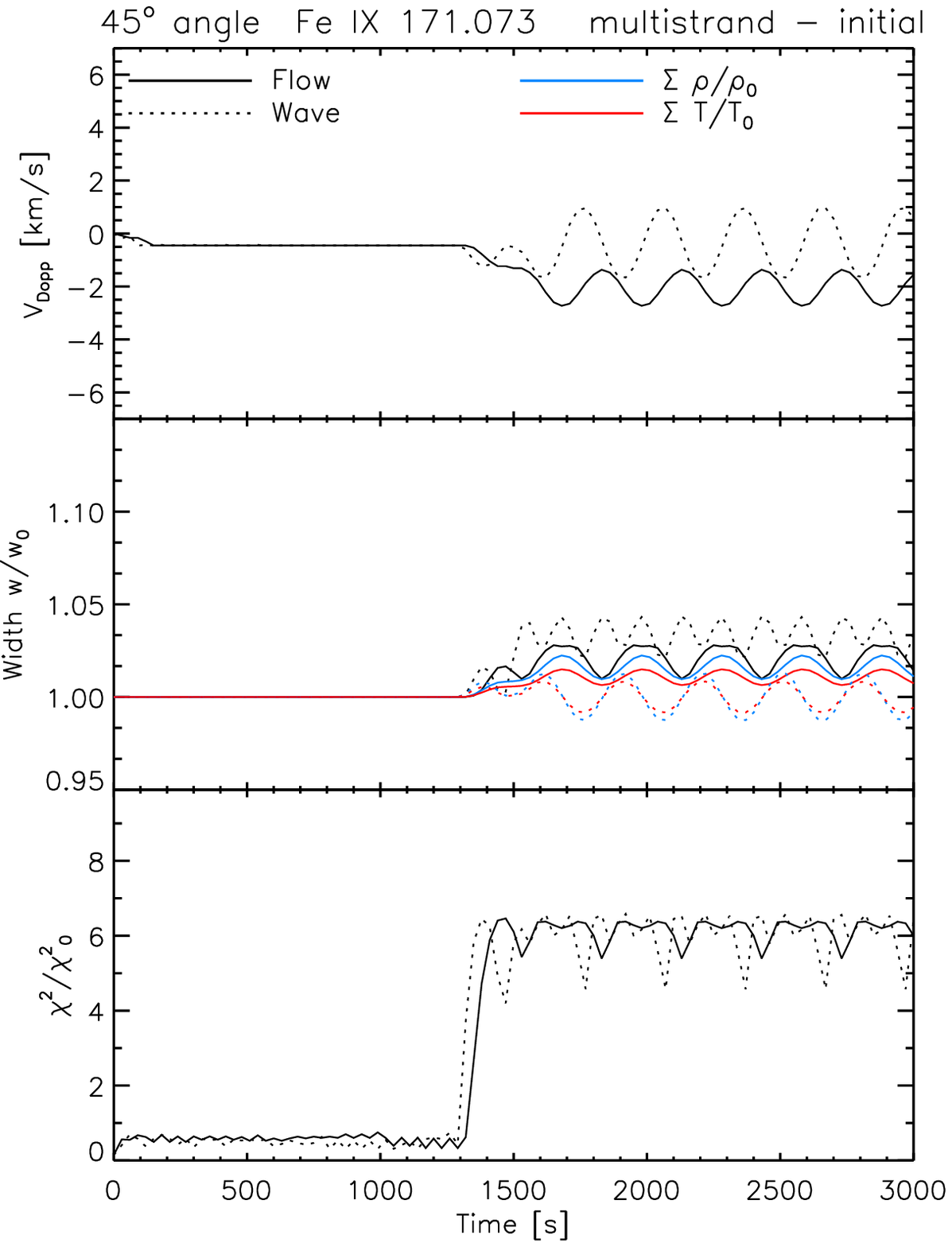}
               \hspace*{-0.03\textwidth}
               \includegraphics[width=0.4\textwidth,clip=]{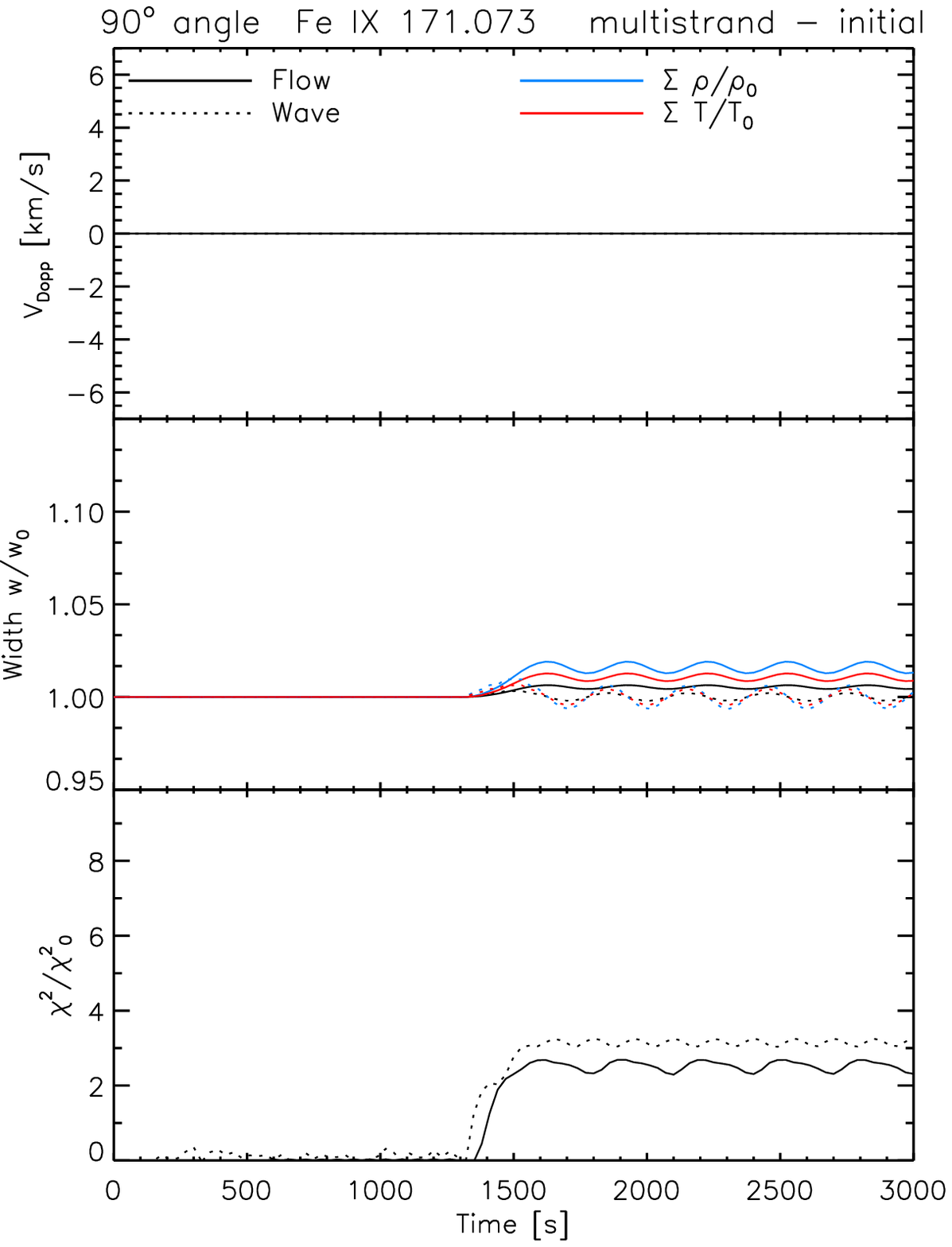}
              }
     \vspace{-0.57\textwidth}  
     \centerline{\bf \small    
      \hspace{0.31 \textwidth}  \color{black}{(c)}
      \hspace{0.31\textwidth}  \color{black}{(d)}
         \hfill}
 \vspace{0.52\textwidth}   
\caption{Time evolution (initial value, multistrand case) of the 171 \AA\  Fe~{\sc{ix}} Doppler velocity (top graphs in four panels), line widths (middle graphs) ,and $\chi^2/\chi_0^2$ (bottom graphs) for tracing rays at angles of (a) $0^\circ$, (b) $15^\circ$, (c) $45^\circ$, and (d) $90^\circ$.}
\label{fig:profiles171_inival_multi}
\end{figure}


Figure~\ref{fig:profiles171_inival_single} shows the temporal evolution of the Doppler velocity (top panels), the line width (middle panels) and the $\chi^2/\chi_0^2$ values (bottom panels) for both the flow (black solid lines) and the wave (black dashed lines) for LOS angles of $\theta=0^\circ, 15^\circ, 45^\circ {\rm ~and~} 90^\circ$. The results shown in Figure~\ref{fig:profiles171_inival_single} correspond to the Fe~{\sc{ix}} 171 {\AA} line. Also overplotted are the summed density and temperature lines (blue and red lines, respectively). These are obtained by summing the density (temperature) perturbations along the LOS, relative to the equilibrium density (temperature) summed over a LOS of the same length ($\int_{{\rm LOS}} \rho dl / \int_{{\rm LOS}} \rho_0 dl$, where $dl$ represents the length along the line-of-sight). 

For $\theta=0^\circ$, all the quantities show a steady change, as these initial value simulations do not reach a steady state before the end of the simulation: the volume of plasma supporting perturbations increases steadily during the simulations. Let us look at the Doppler velocity first. As would be expected for a (periodic) flow travelling towards the observer (and not in a steady state), the flow Doppler velocity keeps increasing.  However, we see that for the wave, there also is a small, net blue shift building up, even though the LOS integrated quantities are averaged over a number of oscillations. This small blue shift results from the fact that for a propagating slow wave, the velocity, density and temperature perturbations are in-phase: as the Doppler velocities are weighted by the density, the integrated LOS values do not add up to zero as one might expect, as ``positive'' perturbations contribute more to the LOS average than the corresponding ``negative" perturbations (see also \citealt{Verwichte2010}). The line-widths (LW) equally keep increasing, with the wave LW increasing more rapidly than the equivalent flow LW. In addition, we see that the wave LWs oscillate with a double frequency. As pointed out by \cite{Verwichte2010} this is caused by the fact that a wave-like perturbation leads to excursions both to the blue and red side of the static line and hence, the resulting total line profile (which consists of the static background plasma plus the perturbed component) will show additional line-broadening for both positive and negative values of the Doppler velocity perturbations, leading to a doubling of the LW oscillation frequency. This also explains why the wave LWs grow more rapidly than the corresponding flow LWs. Although there are some other (physically) real differences in the Doppler velocity and LW evolution of the flow and wave model, these differences are only apparent in a direct comparison and hence cannot be used to observationally distinguish between the two models.

As can be seen from Figure~\ref{fig:rays}(a), a steady state will be reached for LOS angles substantially greater than $0^\circ$ before the end of the simulation. Even though this is an initial value simulation, stopped before the perturbations reach the top boundary, for larger LOS angles a steady state is reached once the perturbations have travelled through the path of the ray. From that time onwards, the simulations are essentially similar to steady-state simulations as on average, no more material is added or perturbed. The larger the LOS angle, the quicker a steady state is reached as the perturbations cross a shorter ray path. The panels of Figure~\ref{fig:profiles171_inival_single} for $\theta=15^\circ$ and $\theta=45^\circ$ indeed show that a steady state is reached quickly, as the Doppler velocity and LWs tend to constant values. 

Finally, we consider the $\theta=90^\circ$ case. Here there is no velocity component aligned with the LOS and hence the Doppler velocities are zero for both the wave and flow model. The doubling of the wave LW frequency is now absent as the Doppler velocity perturbations are zero. The remaining changes in the wave LW are thermal LW changes, associated with the temperature and density perturbations.  However, for $\theta=90^\circ$ the goodness-of-fit measure now shows a doubling in frequency for the wave model which is caused by the fact that the LW oscillates around its equilibrium value: both a maximum and a minimum in the wave LW corresponds to a maximum in $\chi^2$. This is not the case for the flow, where the LW minima actually correspond to the equilibrium values: the perturbations in temperature and density for the flow are only ever positive relative to the equilibrium values.

\subsection{Multistrand Wave and Flow}

Figure~\ref{fig:profiles171_inival_multi} shows the results for the initial value, multistrand simulation. Here five strands of different widths are represented, supporting oscillations with the same frequency but slightly out of phase in time. Again the simulation is stopped before any of the perturbations reach the upper boundary. For $\theta=0^\circ$ we get exactly the same results as in the single strand case described above, as the LOS ray only crosses one of the strands. As before, we again see a doubling of the frequency for the wave LW, which is not present in the corresponding Doppler velocity perturbations. For inclined strands ($\theta=15^\circ$ and $\theta=45^\circ$) we see that the amplitude of the wave LW oscillations is no longer constant, due to the complex addition of out-of-phase perturbations along the LOS. This is also obvious from the $\chi^2$ values, which no longer show harmonic oscillations. The flow LWs, on the other hand, do appear to keep their constant amplitudes but have lost their sinusoidal shape. We also see some evidence of frequency doubling for the flow $\chi^2$ measurements although the effect is probably too small to be observable.

\begin{figure}[t]
\centerline{\hspace*{0.015\textwidth}
               \includegraphics[width=0.4\textwidth,clip=]{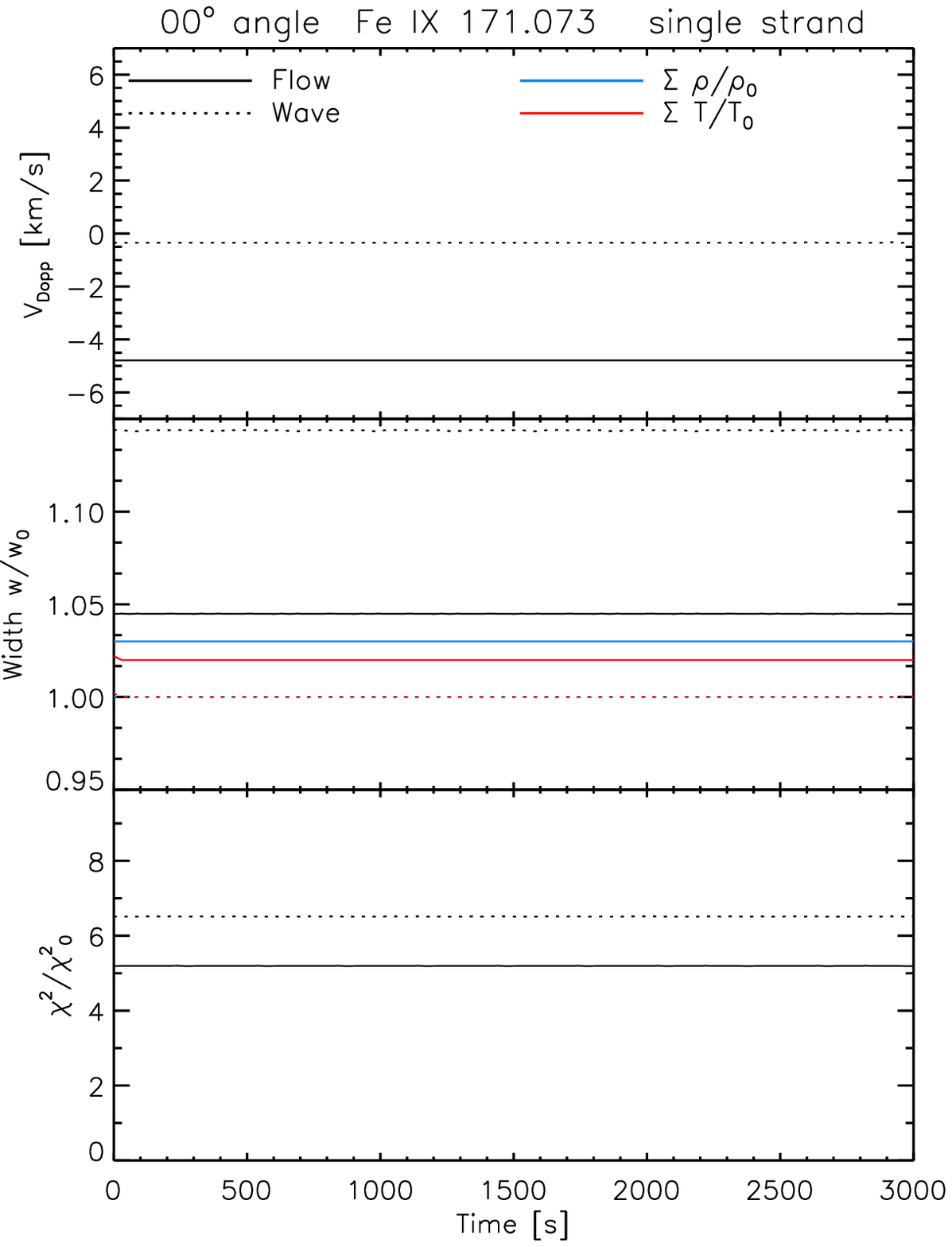}
               \hspace*{-0.03\textwidth}
               \includegraphics[width=0.4\textwidth,clip=]{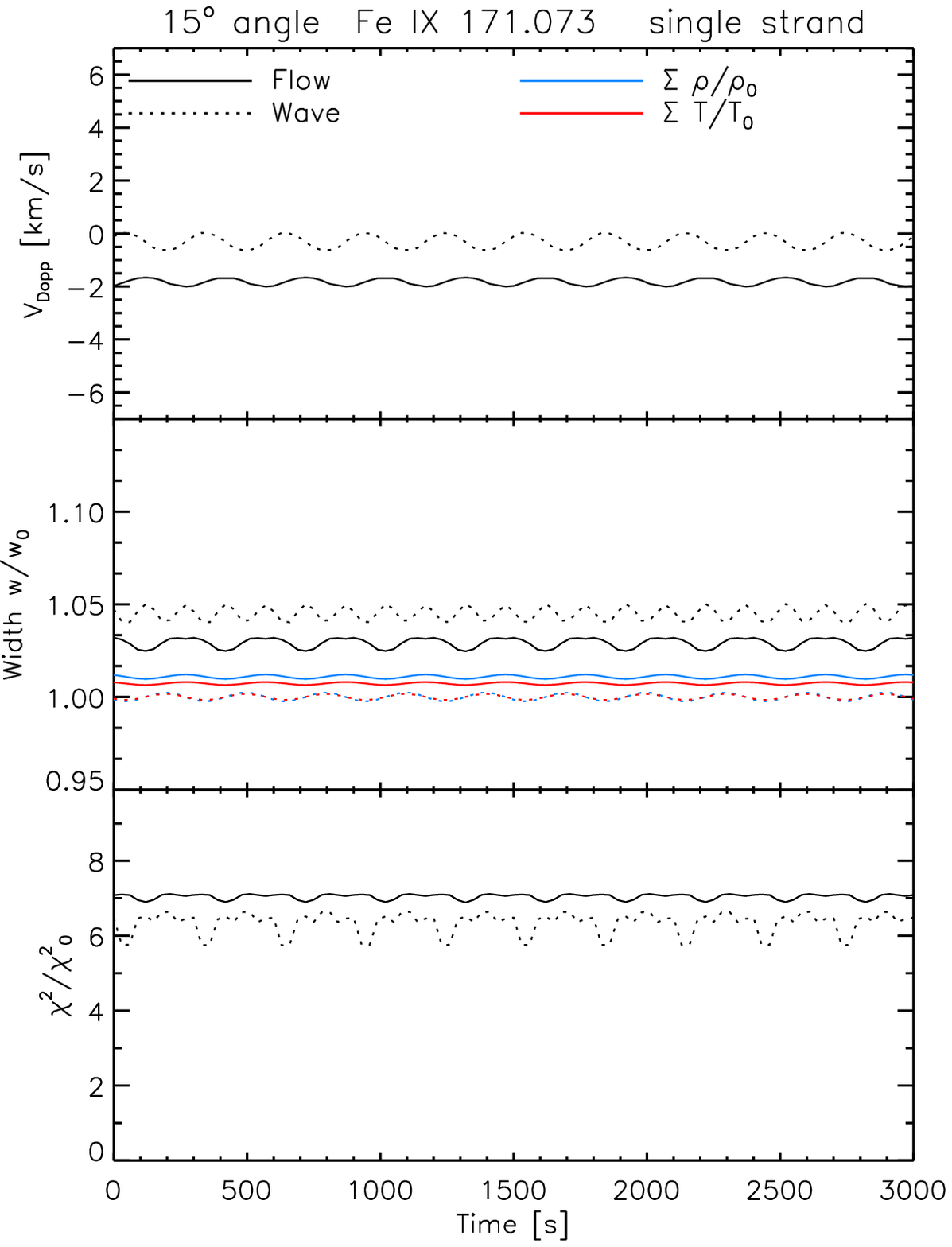}}
 \vspace{-0.57\textwidth}  
     \centerline{\bf \small    
      \hspace{0.31 \textwidth}  \color{black}{(a)}
      \hspace{0.31\textwidth}  \color{black}{(b)}
         \hfill}
  \vspace{0.55\textwidth}    
\centerline{\hspace*{0.015\textwidth}               
                \includegraphics[width=0.4\textwidth,clip=]{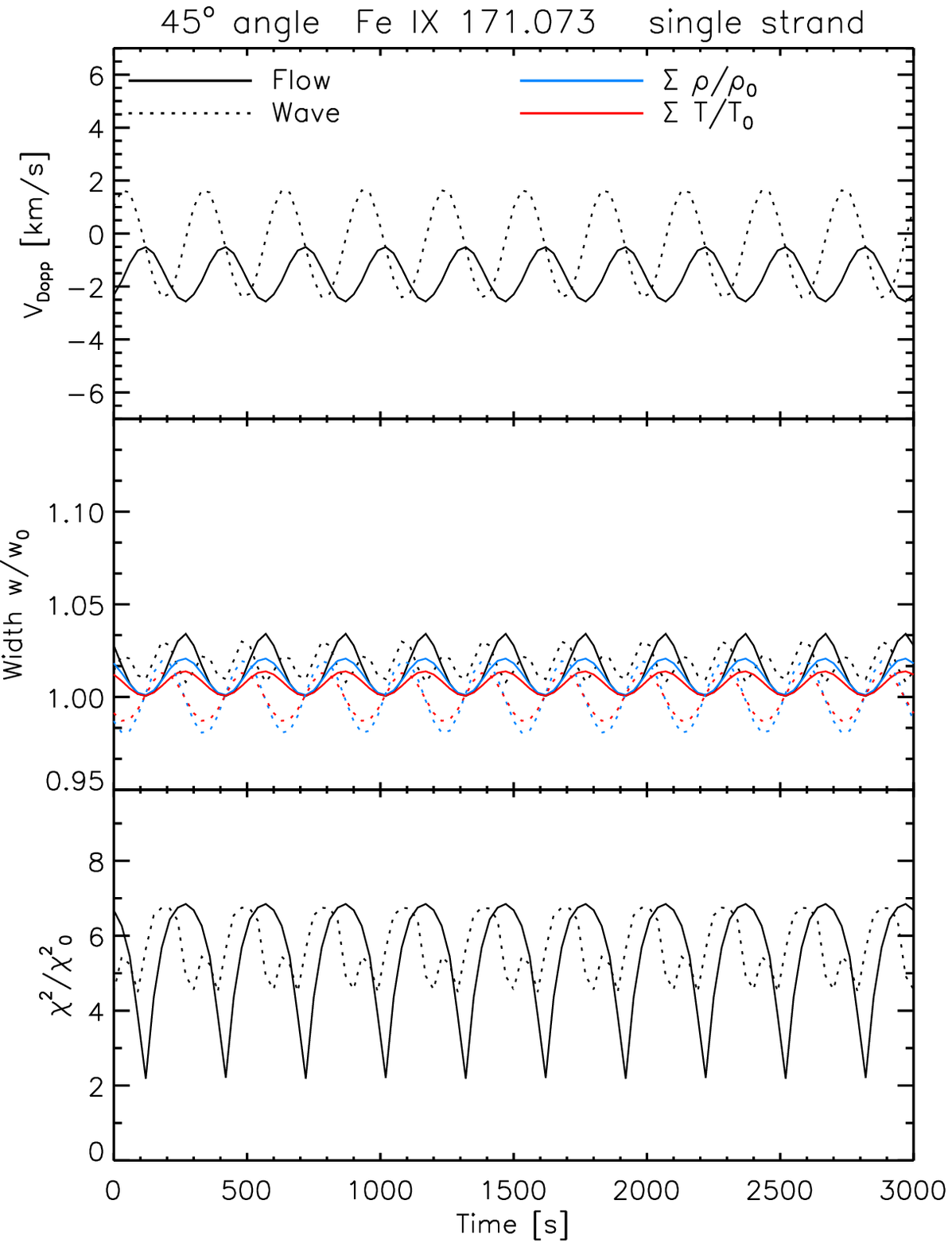}
               \hspace*{-0.03\textwidth}
               \includegraphics[width=0.4\textwidth,clip=]{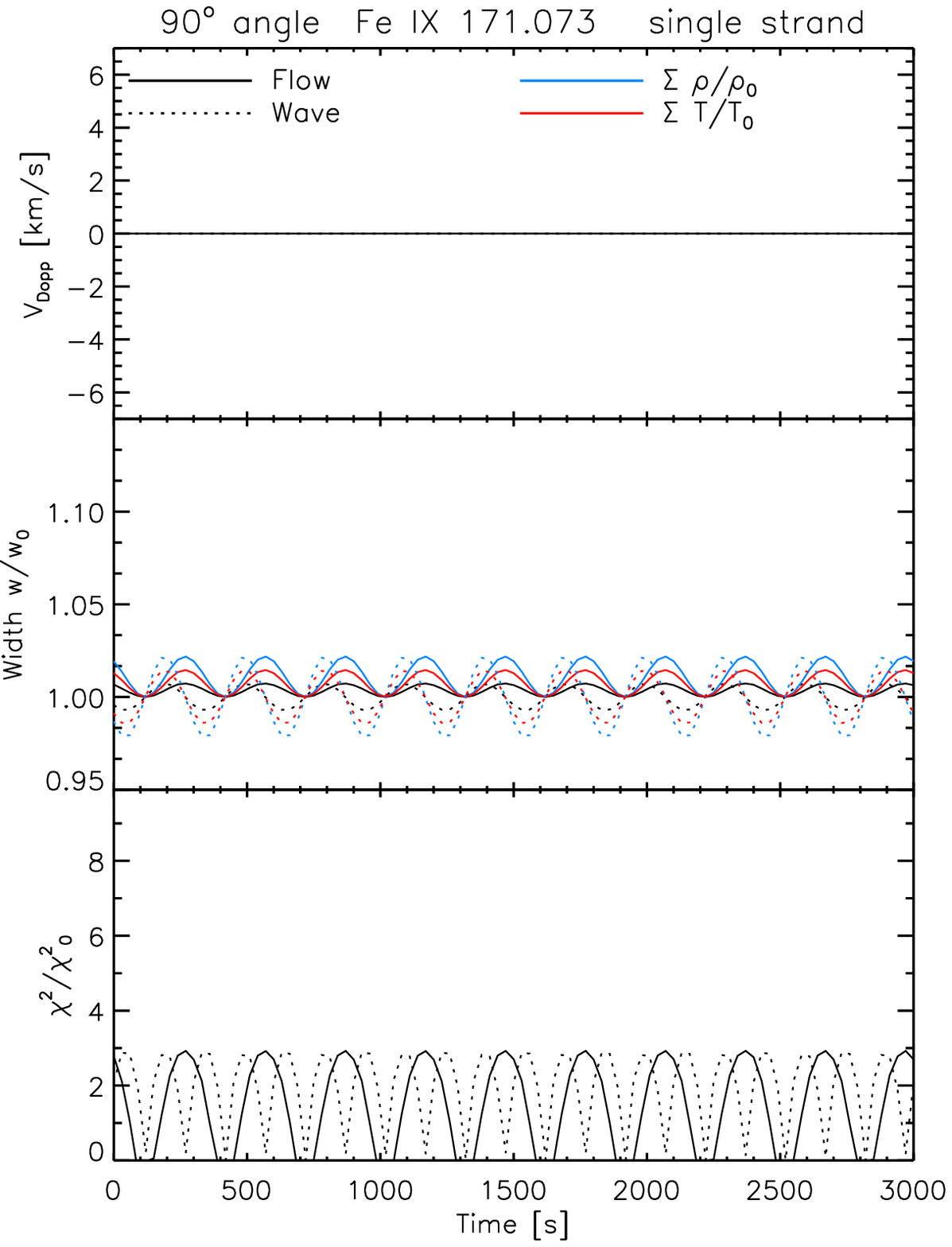}
              }
     \vspace{-0.57\textwidth}  
     \centerline{\bf \small    
      \hspace{0.31 \textwidth}  \color{black}{(c)}
      \hspace{0.31\textwidth}  \color{black}{(d)}
         \hfill}
 \vspace{0.52\textwidth}   
\caption{Time evolution (harmonic, single-strand case) of the 171 \AA\  Fe~{\sc{ix}} Doppler velocity (top graphs in four panels), line widths (middle graphs), and $\chi^2/\chi_0^2$ (bottom graphs) for tracing rays at angles of (a) $0^\circ$, (b) $15^\circ$, (c) $45^\circ$, and (d) $90^\circ$.}
\label{fig:profiles171_harm_single}
\end{figure}


\section{Harmonic Results}\label{sec:results_hmn}

Although the initial-value results described in Section \ref{sec:results_ini} are useful to help understand the behaviour of the wave and flow observational signatures, in practice it is unlikely that observations will show initial-value perturbations (in other words, it is unlikely that we will observe the actual start of a train of perturbations). Hence, we also model `steady-state', harmonic perturbations, {\it i.e.}~the same perturbation (constant period) is travelling through the domain at all times. An example snapshot taken from the multistrand simulations is shown in Figure~\ref{fig:rays}(b). As the numerical simulations used in this paper are computationally cheap, reflection from the top boundary is prevented by stopping the simulations before they reach the top boundary of a much larger box and then only considering the lower half of the box when forward modelling the numerical results.

\subsection{Single Strand Wave and Flow}

As expected for a steady state, all quantities now reach (on average) constant values (see Figure~\ref{fig:profiles171_harm_single}). For $\theta=0^\circ$ we notice that there do not appear to be oscillations in any of the observable quantities. There is a small, residual blue shift for the wave, as discussed earlier and a larger blueshift for the flow but these are both constant values. Observationally, constant values are difficult to measure directly as typical solar spectrometers lack absolute calibration and hence this difference is not a useful signature to distinguish between the periodic flow and wave models. The fact that there are no oscillations is because there is an exact number of periods in the numerical domain in the vertical ($y$) direction ({\it i.e.}~the direction of propagation). Hence, for a loop that is exactly aligned with the LOS, no oscillations would be visible in this case. However, this is a somewhat unlikely scenario and in reality, it is more likely that a small residual oscillation would remain (in other words, it is unlikely that the oscillations would exactly cancel each other out along the LOS). We also note that our oscillations have a constant amplitude whereas in reality, the amplitudes of the propagating disturbances are observed to decay as they travel along coronal loops.

\begin{figure}[t]
\centerline{\hspace*{0.015\textwidth}
               \includegraphics[width=0.4\textwidth,clip=]{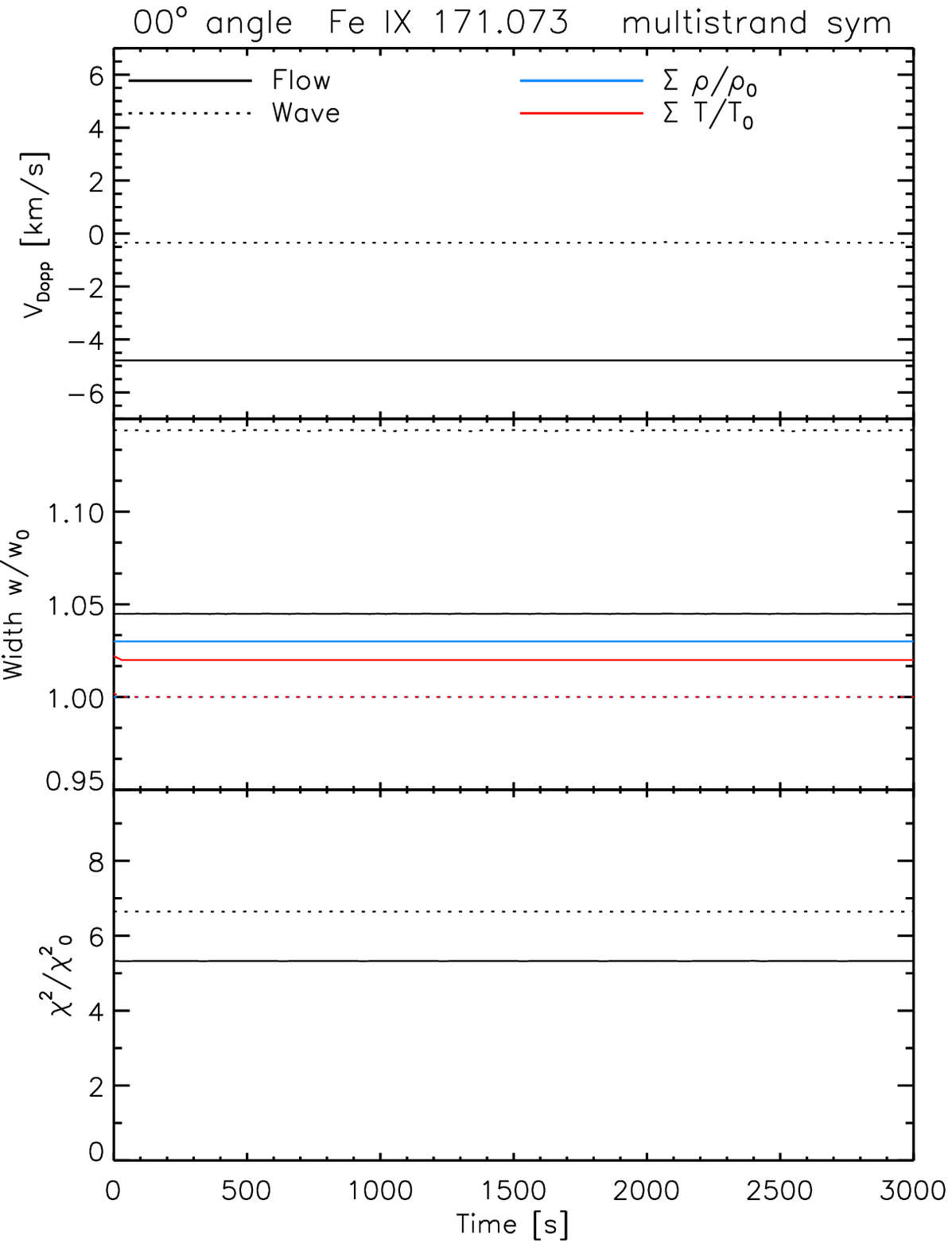}
               \hspace*{-0.03\textwidth}
               \includegraphics[width=0.4\textwidth,clip=]{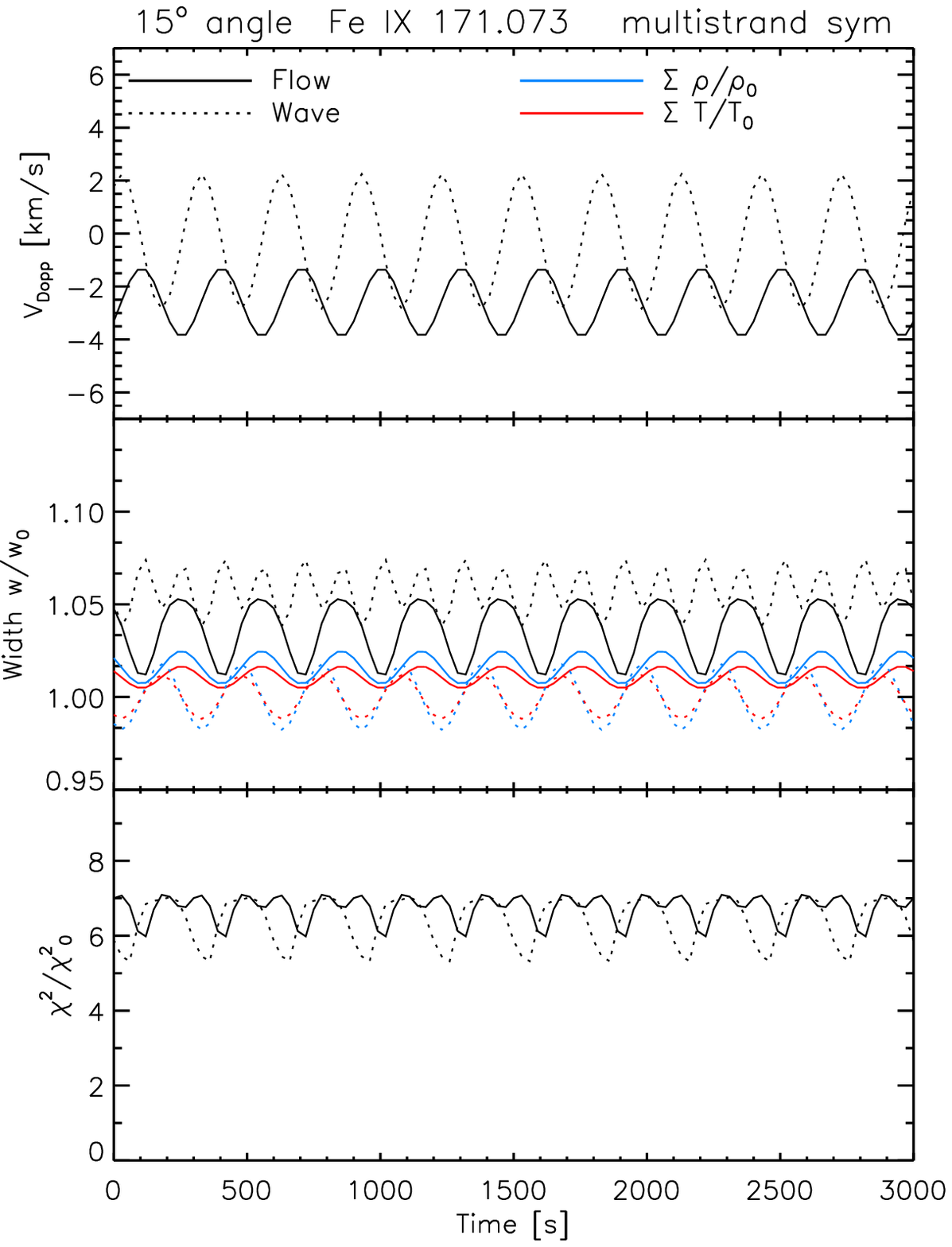}}
 \vspace{-0.57\textwidth}  
     \centerline{\bf \small    
      \hspace{0.31 \textwidth}  \color{black}{(a)}
      \hspace{0.31\textwidth}  \color{black}{(b)}
         \hfill}
  \vspace{0.55\textwidth}    
\centerline{\hspace*{0.015\textwidth}               
                \includegraphics[width=0.4\textwidth,clip=]{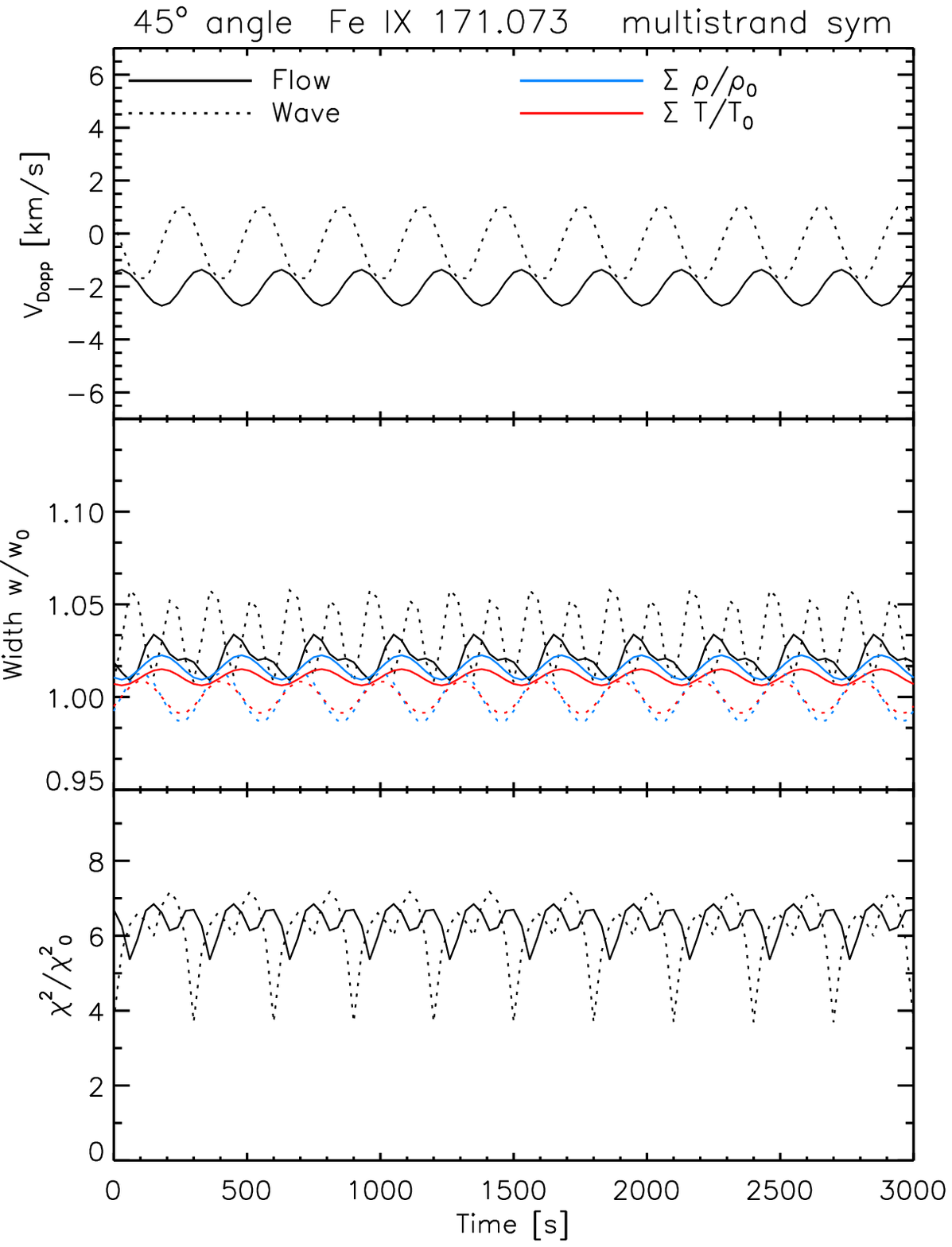}
               \hspace*{-0.03\textwidth}
               \includegraphics[width=0.4\textwidth,clip=]{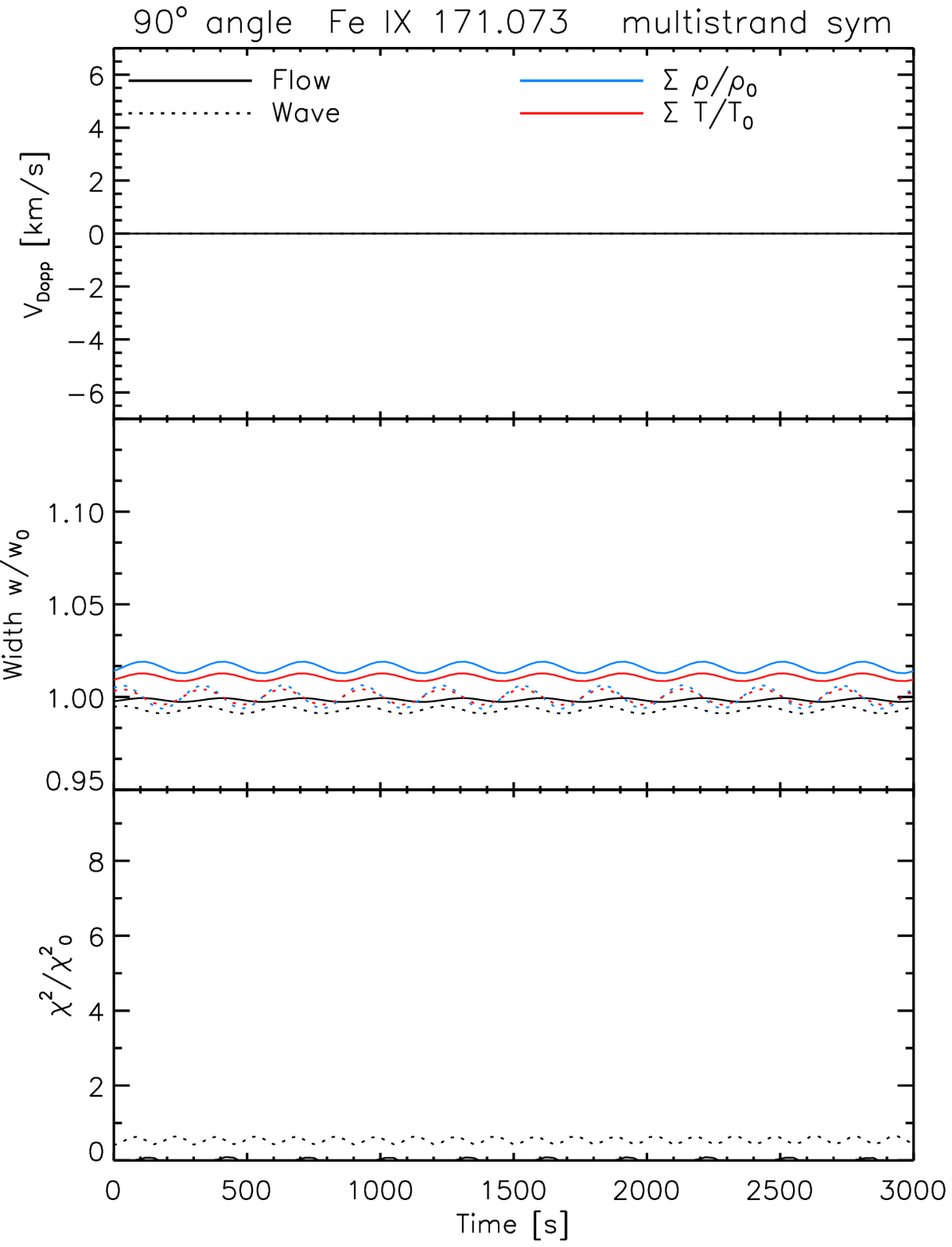}
              }
     \vspace{-0.57\textwidth}  
     \centerline{\bf \small    
      \hspace{0.31 \textwidth}  \color{black}{(c)}
      \hspace{0.31\textwidth}  \color{black}{(d)}
         \hfill}
 \vspace{0.52\textwidth}   
\caption{Time evolution (symmetric harmonic, multistrand case) of the 171 \AA\  Fe~{\sc{ix}} Doppler velocity (top graphs in four panels), line widths (middle graphs), and $\chi^2/\chi_0^2$ (bottom graphs) for tracing rays at angles of (a) $0^\circ$, (b) $15^\circ$, (c) $45^\circ$, and (d) $90^\circ$.}
\label{fig:profiles171_sym_multi}
\end{figure}


\begin{figure}[t]
\centerline{\hspace*{0.015\textwidth}
               \includegraphics[width=0.4\textwidth,clip=]{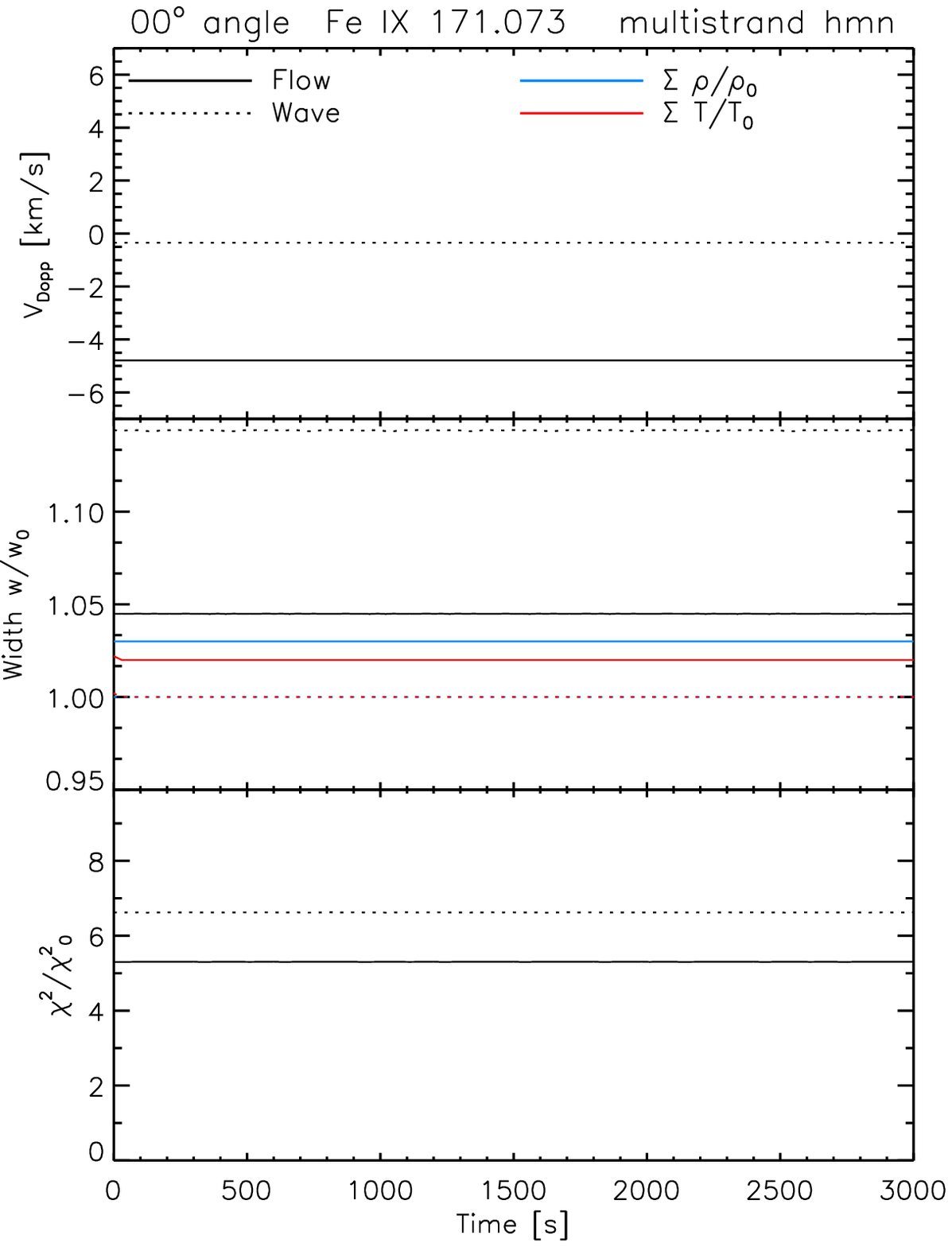}
               \hspace*{-0.03\textwidth}
               \includegraphics[width=0.4\textwidth,clip=]{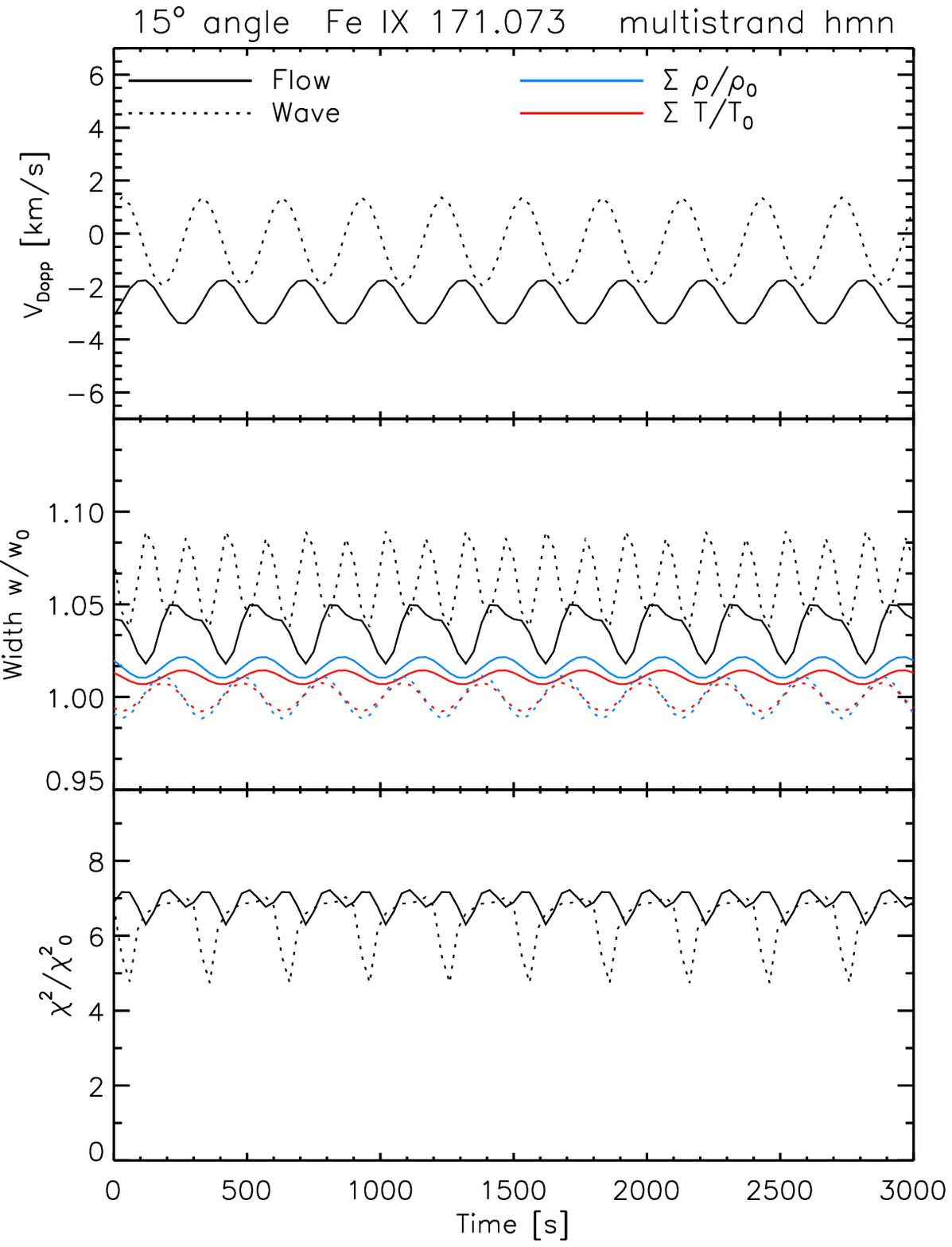}}
 \vspace{-0.57\textwidth}  
     \centerline{\bf \small    
      \hspace{0.31 \textwidth}  \color{black}{(a)}
      \hspace{0.31\textwidth}  \color{black}{(b)}
         \hfill}
  \vspace{0.55\textwidth}    
\centerline{\hspace*{0.015\textwidth}               
                \includegraphics[width=0.4\textwidth,clip=]{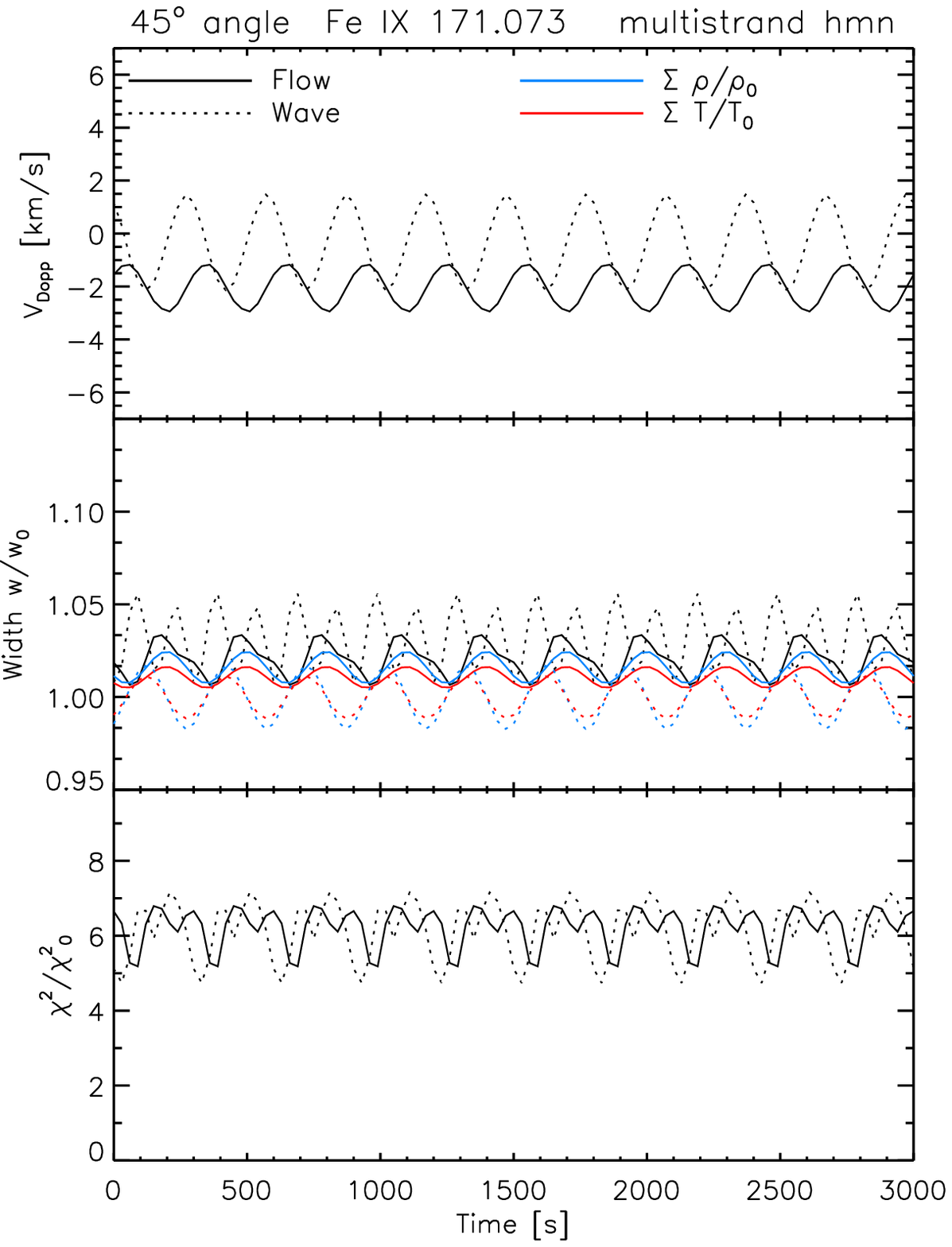}
               \hspace*{-0.03\textwidth}
               \includegraphics[width=0.4\textwidth,clip=]{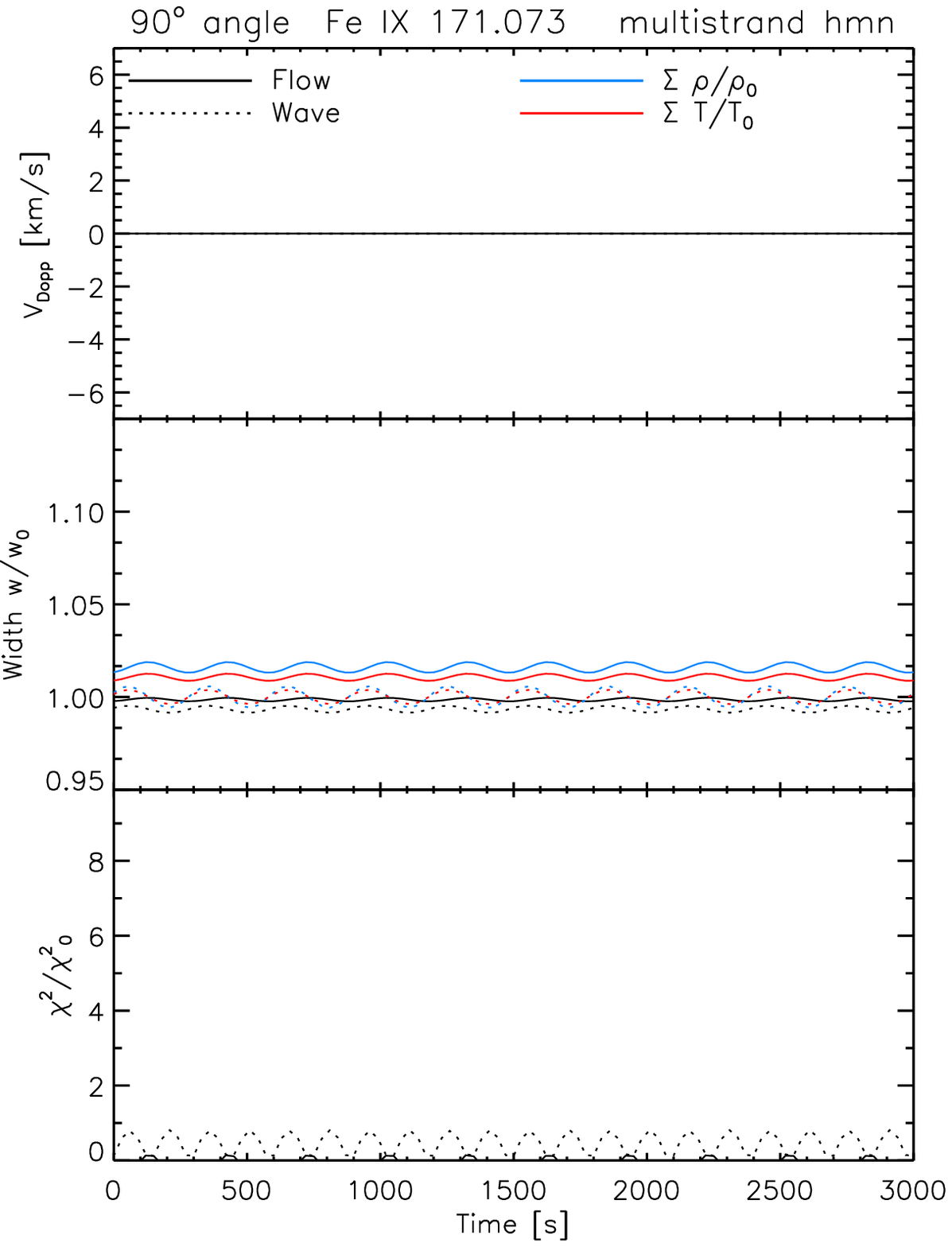}
              }
     \vspace{-0.57\textwidth}  
     \centerline{\bf \small    
      \hspace{0.31 \textwidth}  \color{black}{(c)}
      \hspace{0.31\textwidth}  \color{black}{(d)}
         \hfill}
 \vspace{0.52\textwidth}   
\caption{Time evolution (harmonic, multistrand case) of the 171 \AA\  Fe~{\sc{ix}} Doppler velocity (top graphs in four panels), line widths (middle graphs), and $\chi^2/\chi_0^2$ (bottom graphs) for tracing rays at angles of (a) $0^\circ$, (b) $15^\circ$, (c) $45^\circ$, and (d) $90^\circ$.}
\label{fig:profiles171_harm_multi}
\end{figure}


As expected from the earlier results, the LW associated with the wave is higher than the flow LW and this remains the case for shallow LOS angles. For example, the wave LWs are still larger than the flow ones for $\theta=15^\circ$ but this is no longer the case for $\theta=45^\circ$. For the intermediate angles ($\theta=15^\circ, 45^\circ$) we again see the doubling of the frequency for the LWs associated with the wave. For $\theta=90^\circ$, the results are exactly the same as for the initial value simulations, as expected (as the ray crosses the loop perpendicularly, the `observer' would see exactly the same signal as soon as the first full period of the perturbations has passed the ray in both cases).

\subsection{Multistrand Wave and Flow}

For the steady state simulations, two different versions of the multistrand model were considered, namely one where all the phase differences between the individual strands are exactly the same (Figure~\ref{fig:profiles171_sym_multi} - referred to as the `symmetric' multistrand case) and one where they are different  (Figure~\ref{fig:profiles171_harm_multi}). Generally the results are similar to the ones described earlier for the single strand or the initial value simulations. Note that the multistrand, initial value simulation described in Section \ref{sec:results_ini} also has symmetric phase differences but slightly different ones from the ones used here. The position and width of the strands and the period of the perturbations have been kept constant. These small differences in the phases of the initial value simulations and the steady-state simulations have a small effect on the observational signatures as can be seen from comparing Figures~\ref{fig:profiles171_inival_multi} and \ref{fig:profiles171_sym_multi}. However, none of these differences lead to observational signatures that could definitively distinguish between periodic flows or slow magneto acoustic waves. Similarly, there are small differences between the symmetric, multistrand steady-state simulation and the non-symmetric case (compare Figures~\ref{fig:profiles171_sym_multi} and \ref{fig:profiles171_harm_multi}) but none that would be observationally distinguishable, apart from the frequency doubling in the wave LWs.

\section{Fe~{\sc{xii}} Results}\label{sec:results_fe12}

\begin{figure}[t]
\centerline{\hspace*{0.015\textwidth}
               \includegraphics[width=0.4\textwidth,clip=]{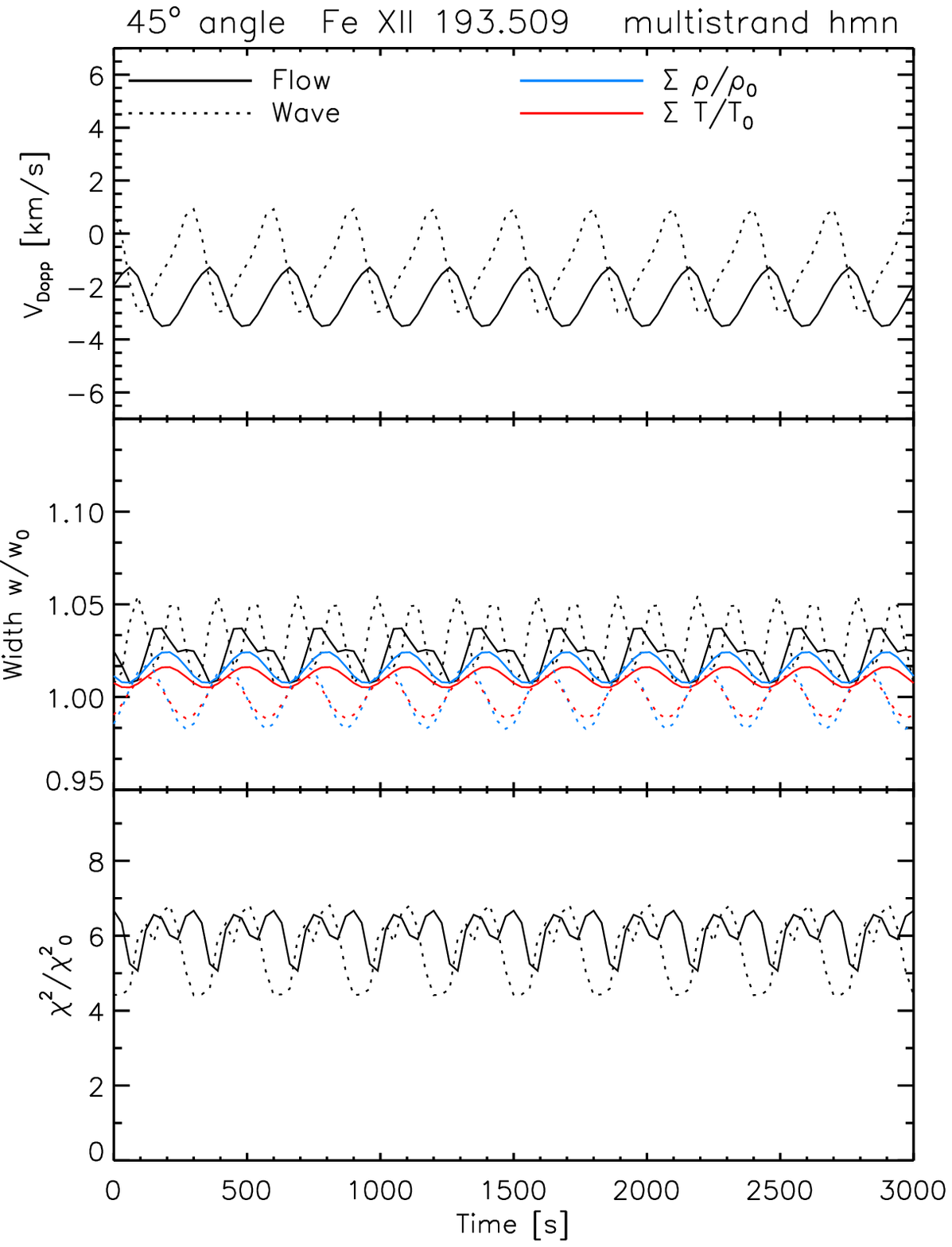}
               \hspace*{-0.03\textwidth}
               \includegraphics[width=0.4\textwidth,clip=]{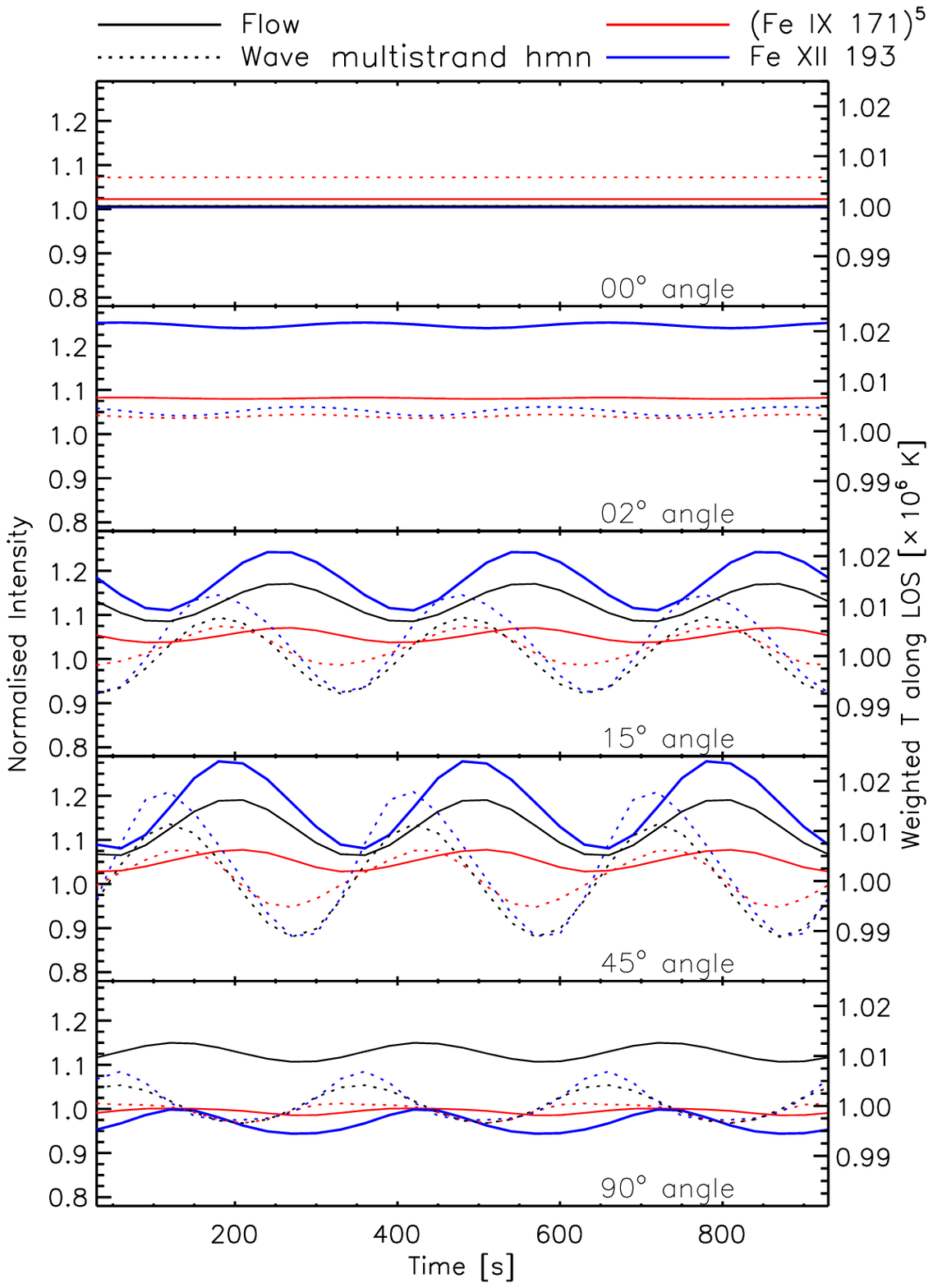}}
 \vspace{-0.57\textwidth}  
     \centerline{\bf \small    
      \hspace{0.31 \textwidth}  \color{black}{(a)}
      \hspace{0.31\textwidth}  \color{black}{(b)}
         \hfill}
   \vspace{0.52\textwidth}   
\caption{(a) Time evolution (harmonic, multistrand case) of the 193 \AA\  Fe~{\sc{xii}} Doppler velocity (top panel), line widths (middle panel) and $\chi^2/\chi_0^2$ (bottom panel) for $\theta=45^\circ$. (b) The observed intensities along different LOS angles for Fe~{\sc{ix}} (red) and Fe~{\sc{xii}} (blue).}
\label{fig:FeXII_ev}
\end{figure}

\begin{figure}[t]
\centerline{\hspace*{0.015\textwidth}
               \includegraphics[width=0.8\textwidth,clip=]{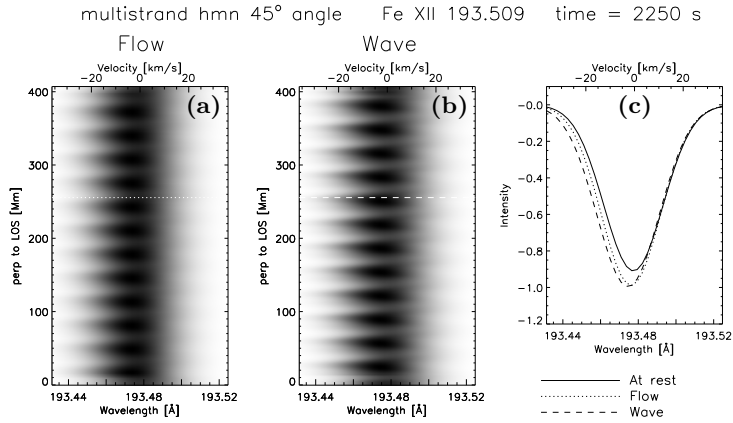}}
 \vspace{-0.375\textwidth}  
     \centerline{\bf \small    
      \hspace{0.295 \textwidth}  \color{black}{(a)}
      \hspace{0.205\textwidth}  \color{black}{(b)}
      \hspace{0.14\textwidth}  \color{black}{(c)}
         \hfill}
   \vspace{0.32\textwidth}   
\caption{A (reverse colour) snapshot (harmonic, multistrand, $45^\circ$ at $t=2250$ s) of the Fe~{\sc{xii}} spectral line as a function of distance perpendicular to the LOS for (a) the flow and (b) the wave. Panel (c) shows a horizontal cut through these snapshots.}
\label{fig:FeXII_snapshot}
\end{figure}

The results described so far all applied to observational signatures in the Fe~{\sc{ix}} 171 {\AA} line. We now briefly look at the Fe~{\sc{xii}} 193 {\AA} line. Overall, the results for Fe~{\sc{xii}} are very similar to the Fe~{\sc{ix}} results in the sense that there are of course difference between the wave and flow results, but none that could be used to observationally differentiate between the quasi-periodic flow or wave interpretations. However, in the multistrand, steady state simulations, an interesting steepening of both the flow and wave Doppler velocities is found at intermediate LOS angles ($\theta = 15^\circ \mbox{and } 45^\circ$), as shown, for example, in Figure~\ref{fig:FeXII_ev}(a) for $\theta=45^\circ$. Excursions to the blue wing appear to happen very rapidly, giving the Doppler velocities a sawtooth appearance reminiscent of shocks, even though there are no actual (physical) shocks present in our numerical simulations. Exactly the same forward-modelled simulations in the Fe~{\sc{ix}} 171 {\AA} line (Figure~\ref{fig:profiles171_harm_multi}©) show no apparent steepening at all, confirming this is entirely an observational effect. The sawtooth pattern can also be seen clearly in Figure~\ref{fig:FeXII_snapshot}, which shows a snapshot of the 193 {\AA} emission for both the flow and the wave as a function of wavelength and distance perpendicular to the LOS.

\begin{figure}[t]
\centerline{\hspace*{0.015\textwidth}
               \includegraphics[width=0.95\textwidth,clip=]{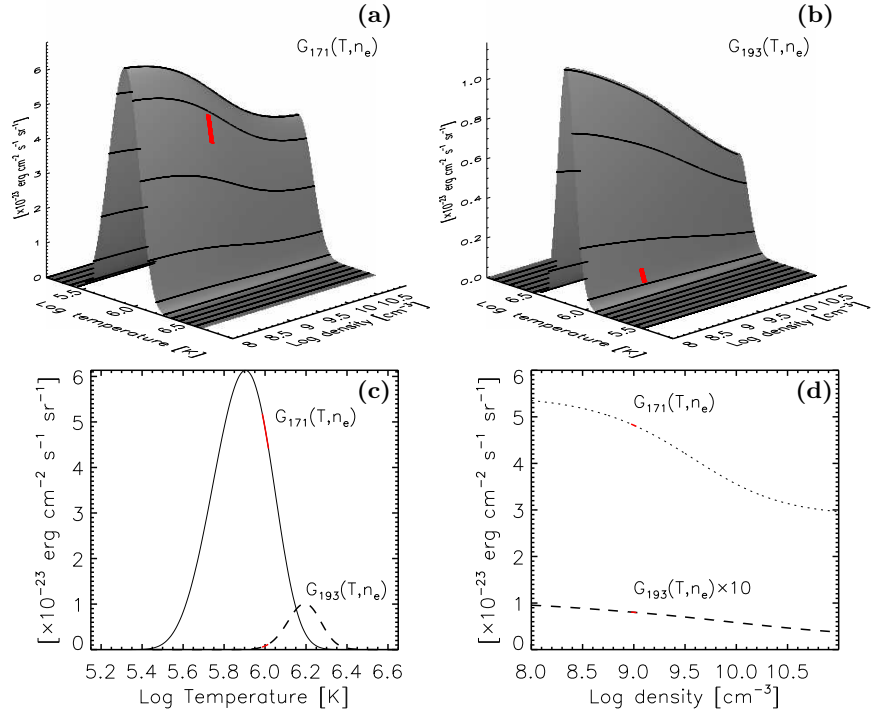}}
 \vspace{-0.8\textwidth}  
     \centerline{\bf \small    
      \hspace{0.405 \textwidth}  \color{black}{(a)}
      \hspace{0.415\textwidth}  \color{black}{(b)}
        \hfill}
 \vspace{0.374\textwidth}  
     \centerline{\bf \small    
      \hspace{0.405\textwidth}  \color{black}{(c)}
      \hspace{0.415\textwidth}  \color{black}{(d)}
         \hfill}
   \vspace{0.35\textwidth}   
\caption{Surface plots of the emission function $G(T,n)$ for (a) 171 \AA\ Fe~{\sc{ix}} and (b) 193 \AA\ Fe~{\sc{xii}} as a function of (log) density and temperature. Note that the temperature axis has been reversed on the $G_{193}(T,n)$ surface plot. The bottom panels show slices through the domain at fixed density (c) and temperature (d). The relevant sections of $G_{171}$ and $G_{193}$ for the temperature and density ranges used in the models are indicated in red.}
\label{fig:G}
\end{figure}

Contributions along the LOS are weighted by $n^2 \times G(T,n)$, where $G(T,n)$ is the plasma response function (see {\it e.g.} \citealt{Antolin2013}). The difference between the 171 and 193 {\AA} lines can be explained by looking in detail at the plasma response function $G(T,n)$ (Figure~\ref{fig:G}). Although this function depends both on temperature and density, for the range of temperatures and densities considered here, the dependence on density is very small and can be ignored (see cross section as a function of density shown in Figure~\ref{fig:G}). The dependence of $G(T,n)$ on temperature is very different for the 171 and 193 {\AA} lines (note that the temperature axis has been reversed on the $G_{193}(T,n)$ surface plot). For 171 {\AA}, $G_{171}(T,n)$ decreases for increasing temperatures (as we are on the `downhill' side of the $G_{171}(T,n)$ function for our chosen background temperature). For 193 {\AA} however, $G_{193}(T,n)$  increases with temperature. Maxima in temperature (and density) are in phase with the minima in the Doppler velocity ({\it i.e.}~the largest blue shifts). For 193 {\AA}, an increase in temperature and density leads to a rapid increase in $n^2 \times G_{193}(T,n)$ (as $n$ and $G_{193}(T,n)$ are both increasing), implying a larger weighting of blueshifted regions along the LOS, leading to the asymmetric, sawtooth pattern in the Doppler velocities, where excursions to the blue side happen very rapidly. This rapid increase is absent in the 171 {\AA} line as the increase in $n$ is tempered by the simultaneous decrease in $G_{171}(T,n)$.

Figure~\ref{fig:FeXII_ev}(b) shows the actual intensity that would be observed for a range of LOS angles. For $\theta=45^\circ$ we can indeed see a similar steepening in the Fe~{\sc{xii}} 193 {\AA} intensity, especially for the wave model (red dash line). However, we also see that the corresponding  intensity of the Fe~{\sc{ix}} 171 {\AA} line shows a steepening in the opposite direction although it is not as pronounced. The apparent ``shocks'' are more pronounced in the intensity perturbations associated with the wave model as the associated temperature and density perturbations are larger than in the flow model.

We emphasise again that this erratic ``shock-like" profile is just an apparent effect, caused by the highly non-linear nature of the forward modelling process (see also \citealt{IDM08}) and the particular width and alignment of the strands in this specific model and is in no way related to actual physical shocks. 

\begin{figure}[t]
\centerline{\hspace*{0.015\textwidth}
               \includegraphics[width=0.99\textwidth,clip=]{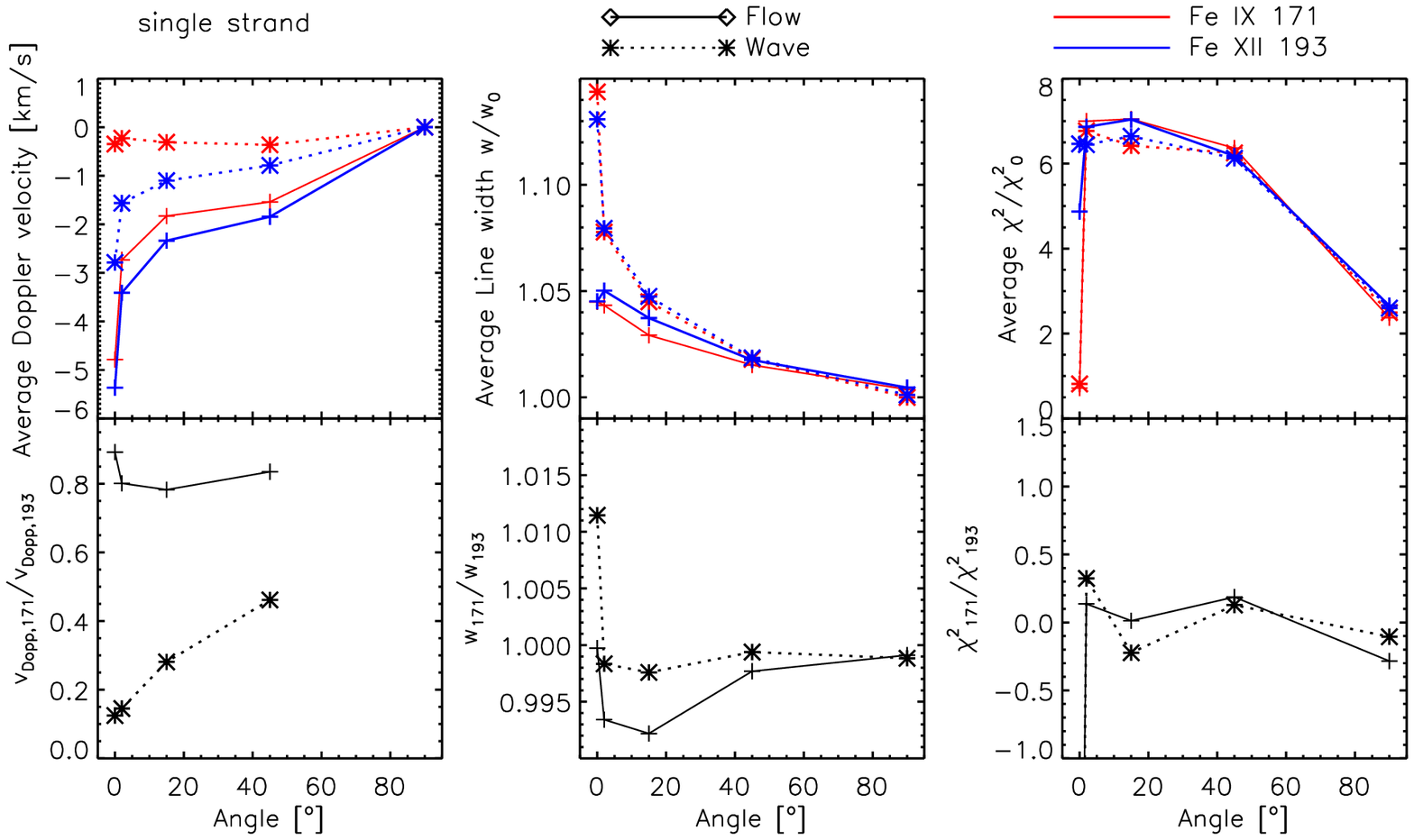}}
 \vspace{-0.635\textwidth}  
     \centerline{\bf \small    
      \hspace{0.18 \textwidth}  \color{black}{(a)}
      \hspace{0.25\textwidth}  \color{black}{(b)}
      \hspace{0.27\textwidth}  \color{black}{(c)}
         \hfill}
   \vspace{0.57\textwidth}   
\caption{Averages of Doppler velocities (a), line widths (b), and $\chi^2/\chi_0^2$ (c) as a function of the LOS angle for the Fe~{\sc{ix}} (red) and Fe~{\sc{xii}} (blue) spectral lines individually (top panels). The ratio of these averages in the two spectral lines is shown in the bottom panels. As before, the results for the flow model are represented by solid lines and the waves model by the dashed lines.}
\label{fig:averages}
\end{figure}


\section{Discussion}\label{sec:disc}

The aim of this paper was to use forward modelling to investigate whether it is actually possible to (observationally) distinguish slow, propagating waves and (quasi-)periodic flows in a basic 2D model. Although there are of course a number of differences in the observational signatures described above, the majority are only identifiable when directly comparing the wave and flow models. However, in practice, a signature needs to be identifiable in isolation and at least in the results presented above, no such signature is readily apparent (especially not when one considers that real observations would be less clear due to the additional effects of the background plasma, the plasma evolution (with height and time) and noise). The doubling of the frequency in the LW perturbations for the waves model could be a potential observational signature but as discussed by \cite{Verwichte2010}, the addition of a modest {\it steady} upflow component would cause the LW to oscillate again with the same period as the other parameters. Hence, these authors remark that the absence of this frequency-doubling in the observed LWs cannot exclude the wave-interpretation of the PCDs as a steady upflow component cannot be excluded (due to the lack of an absolute calibration of the spectrometers). We will comment further on the LW frequency-doubling below.

To try and avoid the need for a comparison between models, we consider the change with LOS angle of the averages of the Doppler velocities, line widths and the $\chi^2/\chi_0^2$ values for both the (single-strand) flow and wave model, for the Fe~{\sc{ix}} 171 {\AA} and Fe~{\sc{xii}} 193 {\AA} lines in Figure~\ref{fig:averages}. We also look at the ratio of these values in the two different spectral lines. A change of LOS angle could observationally be studied by tracking a loop (fan) system showing propagating coronal disturbances over a few days as it rotates across the solar disk whereas ratios of the different quantities in different spectral lines could be obtained using data from {\it e.g.}~{\it Hinode}/{\it EUV Imaging Specrtograph} (EIS; Culhane {\it et al.}, 2007) or {\it Interface Region Imaging Spectrograph} (IRIS; De Pontieu {\it et al.}, 2014). The results are only shown for the single-strand model as those for the two multistrand models are very similar. 

There are a few small differences between the two models. As expected, the Doppler velocities in the flow model remain larger (in absolute values) than for the wave model. The line widths on the other hand appear slightly larger for the wave model but the difference would probably not be observable. We do see however that the LWs for the wave model change very rapidly for shallow LOS angles. In addition, the ratio of the Doppler velocities in the 171 and 193 {\AA} lines is closer to one for the flow model than in the wave model. In other words, Doppler velocities derived from the two spectral lines are closer together for the flow model and further apart for the waves model, again especially at shallow LOS angles.

\begin{figure}[t]
\centerline{\hspace*{0.015\textwidth}
               \includegraphics[width=0.45\textwidth,clip=]{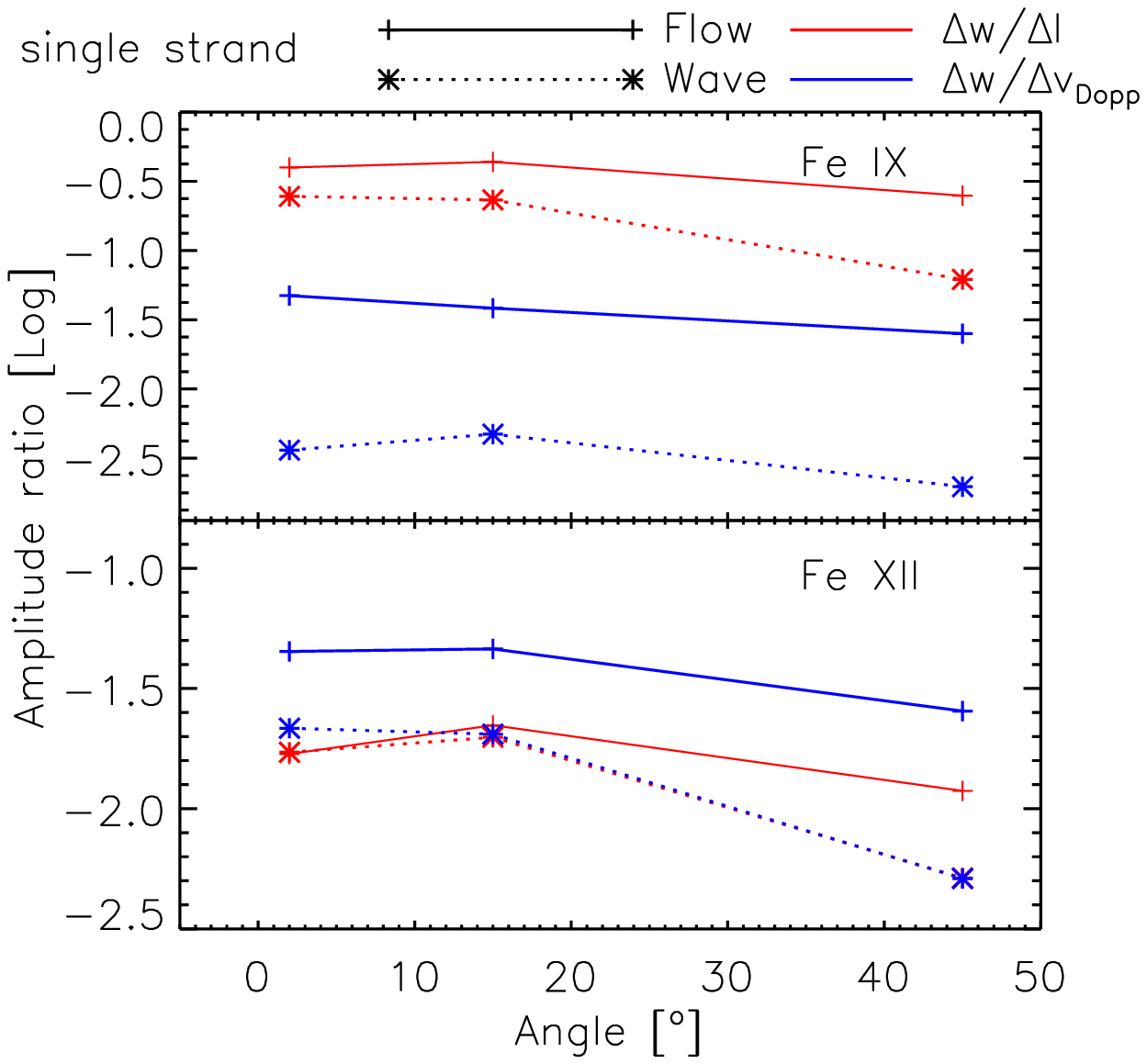}
               \hspace*{-0.01\textwidth}
               \includegraphics[width=0.45\textwidth,clip=]{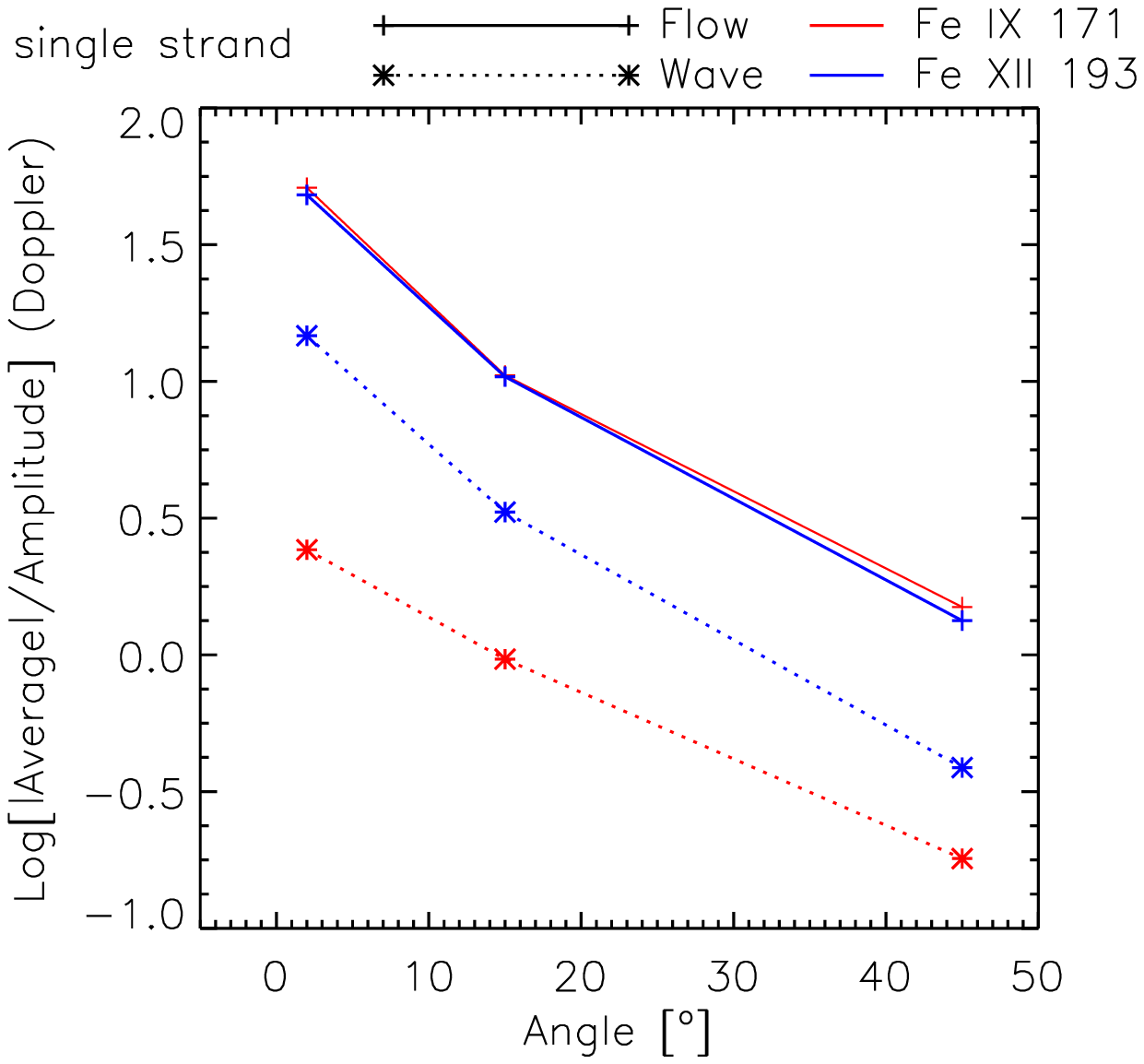}}
 \vspace{-0.48\textwidth}  
     \centerline{\bf \small    
      \hspace{0.28 \textwidth}  \color{black}{(a)}
      \hspace{0.39\textwidth}  \color{black}{(b)}
         \hfill}
   \vspace{0.44\textwidth}   
\caption{(a) Ratio of the amplitudes of the line widths and the intensities ($\Delta w / \Delta I$ - red lines) and the amplitudes of the line widths and the Doppler perturbations ($\Delta w / \Delta v_{{\rm Dopp}}$ - blue lines) as a function of the LOS angle, for Fe~{\sc{ix}} (top panel) and Fe~{\sc{xii}} (bottom panel). (b) Ratio of the average Doppler velocity and the amplitude of the Doppler velocities for Fe~{\sc{ix}} (red) and Fe~{\sc{xii}} (blue).}
\label{fig:amp_ratios}
\end{figure}


Apart from looking at the ratios in different spectral lines, one could also look at the ratios between different amplitudes. Figure \ref{fig:amp_ratios} shows the (logarithm of) the ratios of the amplitudes of the line widths and the intensities ($\Delta w / \Delta I$ - red lines) and the line width amplitudes and the amplitudes of the Doppler perturbations ($\Delta w / \Delta v_{{\rm Dopp}}$ - blue lines) for both the flow and wave models, in the Fe~{\sc{ix}} (top panel) and Fe~{\sc{xii}} (bottom panel) spectral lines. Looking at the bottom panel (Fe~{\sc{xii}} line), the two ratios are essentially the same for the wave model at every angle. However, for the flow model, there is a substantial difference between $\Delta w / \Delta I$ and $\Delta w / \Delta v_{{\rm Dopp}}$ in both spectral graphs. These graphs suggest that $\Delta w / \Delta v_{{\rm Dopp}}$ could potentially be used to distinguish between flows and waves. In Fe~{\sc{ix}}, the ratio $\Delta w / \Delta v_{{\rm Dopp}}$ is substantially bigger for the flow than for the wave (by about an order of magnitude - compare the solid and dashed blue lines in the top panel). For the flow, $\Delta w / \Delta v_{{\rm Dopp}}$ is of the order of $10^{-1.0}$ or $10^{-1.5}$ (so around
0.03-0.1) whereas for the wave model, $\Delta w / \Delta v_{{\rm Dopp}} \approx 10^{-2.5} \approx 0.003$, which is considerably lower. In Fe~{\sc{xii}}, the flow and wave $\Delta w / \Delta v_{{\rm Dopp}}$ ratios are somewhat closer together but still at least a factor of 5 different. Such a large discrepancy may be directly observed, even without the other case as comparison ({\it i.e.}~if one observes a large value for $\Delta w / \Delta v_{{\rm Dopp}}$, the flow scenario might be more likely.) However, the amplitude ratios do depend on the amplitude of the PCDs and properties of the equilibrium state, and the effects of those on the amplitude ratio has not been modelled in this paper. 

Perhaps the most straightforward `usable' signature to distinguish observationally is the simple fact that the average Doppler velocities of the wave model are close to zero. Although there is a net blue shift in the wave model due to the in-phase relationship between the perturbed density and temperature and the velocities, it is generally small. This can also be seen in the top panel of Figure 2 of \cite{Verwichte2010} where the (averaged) wave Doppler velocity is plotted as a function of the wave amplitude. At least for linear perturbations ($v \lsim 60$ km s$^{-1}$  in Figure~2 of \citealt{Verwichte2010}) the average Doppler velocities are smaller than the wave amplitudes.  Hence, one way of `differentiating' a slow wave from a periodic flow could be the interpretation that for a wave model, the amplitude of the Doppler velocity perturbations is larger than the average of the perturbations. Figure \ref{fig:amp_ratios}(b) shows the ratios of the absolute value of the mean of the Doppler velocities to the amplitude of the Doppler velocities. Here, we see that this ratio is indeed larger than one (larger than zero in the figure as the logarithm of the ratio has been plotted) for the flow for all LOS angles. In addition, the ratio is the same in both spectral lines for the flow model. Although we only show the results for the single strand case, the same holds for both multistrand cases. For the wave model, the ratio is smaller than one (less than zero in the logarithm plot) for Fe~{\sc{ix}} apart from at very shallow LOS angles. For Fe~{\sc{xii}} though, the ratio is only less than one for the wave for large LOS angles. However, for both the single strand and the multistrand cases, the ratio in Fe~{\sc{xii}} is substantially bigger than in Fe~{\sc{ix}} at all LOS angles (as opposed to the equal ratios for the flow model) and this could potentially be verified in spectral observations.

\begin{figure}[t]
\centerline{\hspace*{0.015\textwidth}
               \includegraphics[width=0.4\textwidth,clip=]{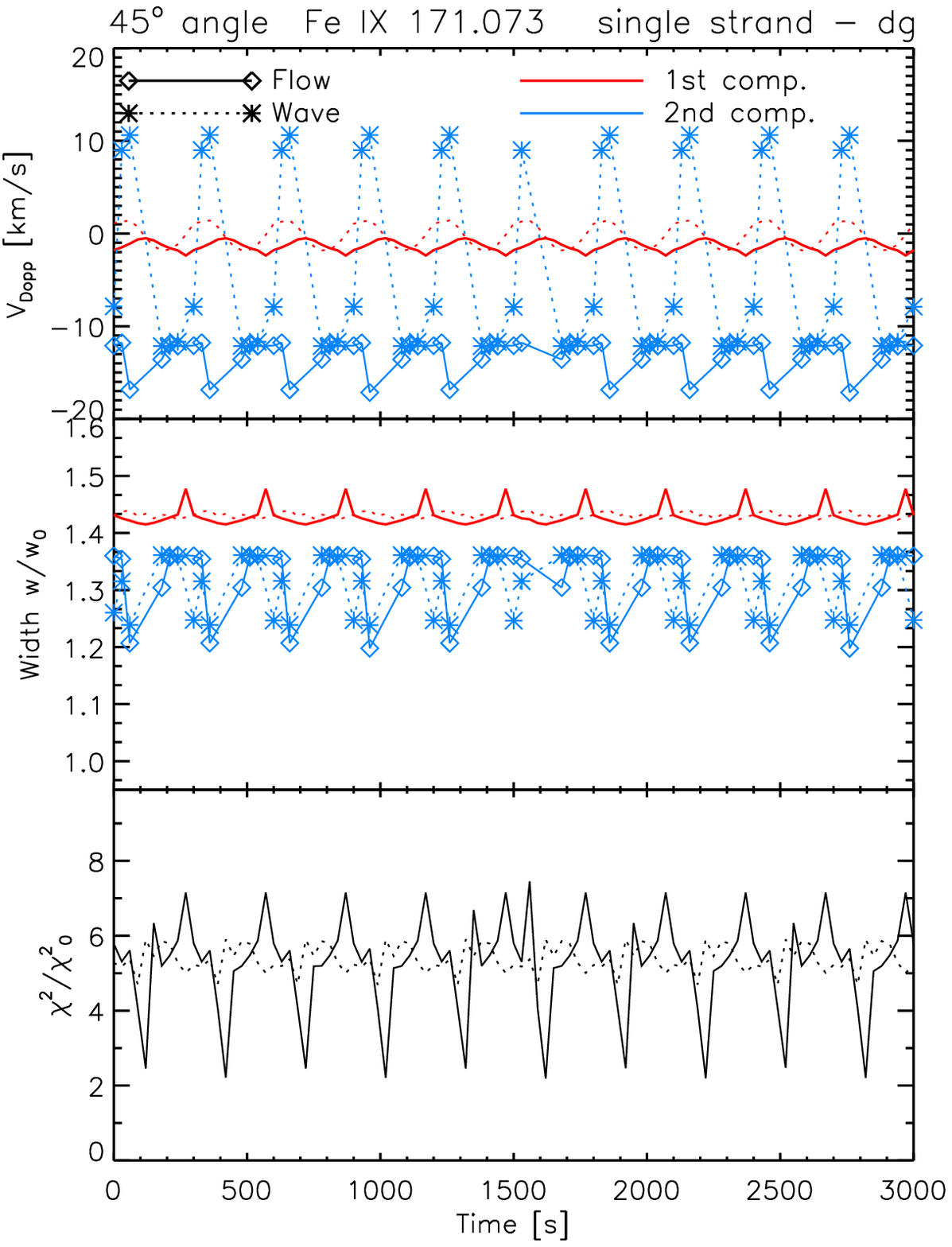}
               \hspace*{-0.03\textwidth}
               \includegraphics[width=0.4\textwidth,clip=]{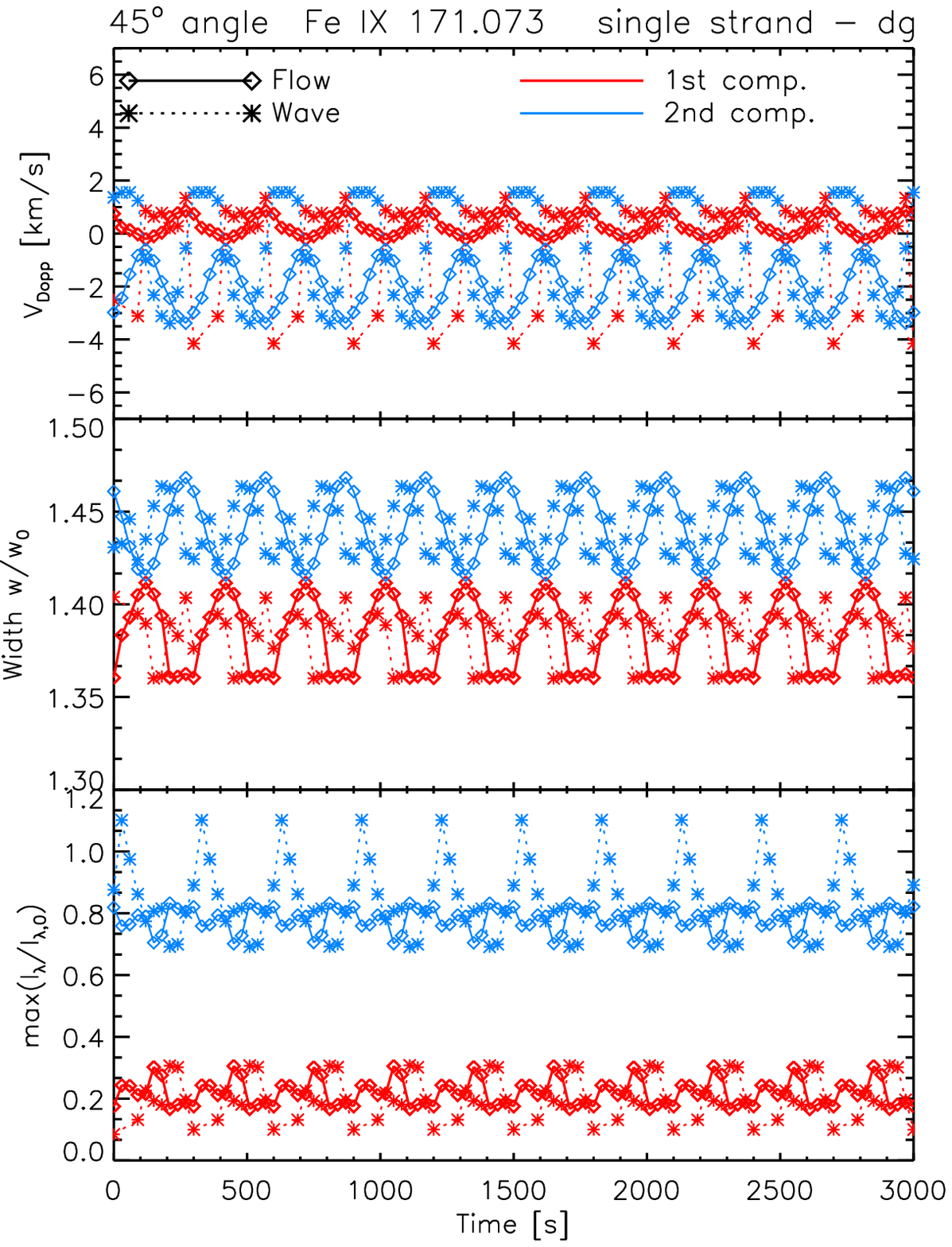}}
 \vspace{-0.57\textwidth}  
     \centerline{\bf \small    
      \hspace{0.31 \textwidth}  \color{black}{(a)}
      \hspace{0.31\textwidth}  \color{black}{(b)}
         \hfill}
   \vspace{0.52\textwidth}   
\caption{Time evolution of the harmonic, single-strand case based on double Gaussian fits (first component in red, second component in blue). Results in panel (a) are obtained without providing an initial guess for the fitting routine, while panel (b) shows results using the single Gaussian fit as an initial guess (see text for more details).  Note that the bottom panel of (b) shows the perturbed intensities rather than the goodness-of-fit measure.}
\label{fig:DGF}
\end{figure}


Theoretically, this difference between the average (or background) Doppler velocities and the perturbation amplitudes could be a feature that would allow distinguishing between the flow and wave interpretation. This is due to the fact that a periodic flow can be decomposed into a steady background flow plus periodic perturbations oscillating around a zero mean, or, in other words, a slow propagating wave, as was pointed out in Section \ref{sec:model}. Equation (\ref{eq:superpos}) shows clearly that the only difference between the flow and wave model is exactly this background value; the periodic flow can be reduced to a slow wave by subtracting the mean of the perturbations ({\it i.e.}~the steady background value).  However, observationally, the difficulty would lie in determining the absolute value of the background: in practice, background values are often subtracted and only the relative amplitudes are studied, which would eliminate this difference between the wave and flow models. Note also that this interpretation implies that the suggestion by \cite{Verwichte2010} to add a modest upflow to the wave model to `avoid' the frequency-doubling in the line widths could essentially transform their wave model into a periodic upflow model (depending on the size of the additional steady upflow component). In other words, if the frequency-doubling of the line-widths is not observed, the interpretation in terms of a quasi-periodic upflow model might be more appropriate. In addition, this interpretation implies the existence of an `intermediate' regime, where the average (background) Doppler velocity and the perturbation amplitudes are similar in size. 

In this study, the periodic flow model we have studied can be decomposed into a steady background flow and wave of the same amplitude due to our particular choice of boundary driving. Of course, various combinations of the background flow amplitude and the perturbation amplitude would lead to different observational signatures but the results presented here show that there will likely only be relative differences between a wave and periodic flow interpretation, rather than (observationally useful) absolute differences.

It is important to point out that the results presented here were obtained by fitting a single Gaussian to the spectral data. This is a crucial difference from, for example, observational results presented by \cite{DePontieu2009} or \cite{DePontieu2010}, who argue that a double Gaussian fit should be used to account for the persistent red-blue asymmetry present in the spectral data (showing an excess in the blue wing when using a single Gaussian fit). In Figure~\ref{fig:DGF} we present two examples of a double Gaussian fit for the steady, single-strand model at $\theta=45^\circ$. Figure~\ref{fig:DGF}(a) is obtained without giving an initial guess to the fitting routine. Figure~\ref{fig:DGF}(b) uses the maximum value, the center position and width of the line at rest as an initial guess for the first component and the center position and line width of the single Gaussian fit as an initial guess for the second component. Figure~\ref{fig:DGF}(a) shows a nearly static first component (red lines) and a secondary component which still oscillates around near-zero for the wave model but around a much larger (blue-shifted) value for the flow. In fact, for the flow, the secondary component appears to significantly overestimate the velocity amplitudes, oscillating around a value of about 15 km s$^{-1}$  whereas the model velocities had maximum values just below 10 km s$^{-1}$  (see Figure~\ref{fig:model}(a)). However, one has to keep in mind that the single Gaussian fit significantly {\it underestimated} the velocity values (see Figure~\ref{fig:profiles171_harm_single}), more so than the overestimate associated with this double Gaussian fit. Hence, one could argue that the double Gaussian fit is actually the better result. In addition, the (small) oscillation in the first component is out of phase with the secondary component for the flow model. When an initial guess is provided (Figure~\ref{fig:DGF}(b)), the flow still shows a first component with only small oscillations around zero and a (now underestimated) blue-shifted second component (again out of phase with the first component). However, the wave model (dashed lines) shows very different behaviour: the first and second components are both oscillating, with similar amplitudes but out of phase. The corresponding LWs show relatively regular behaviour for the flow model, with the first and second components oscillating out of phase, and again show some evidence of frequency doubling for the wave model.

From the first double Gaussian example, we can see that for the wave model, both the first and secondary component would show a similar mean (close to zero in our simple model) whereas for the periodic flows, the mean of the secondary component differs substantially from the static (background) component. This essentially leads to the same conclusion as the single Gaussian fits presented in this paper: quasi-periodic flows are characterised by perturbed Doppler velocities which have amplitudes less than their mean whereas for the slow, propagating waves, the perturbations amplitudes are bigger than the mean (which will be close to zero). However, the second double Gaussian fit shows worryingly different results, indicating the fitting is very sensitive to the initial guess provided, and this could also be the case for low signal-to-noise (`noisy') observations. Hence, one would need to provide appropriately (physically) justified initial guesses for the fitting routine, but it is hard to see how this can be done without pre-determining the physical model.  The double Gaussian fitting used by \cite{DePontieu2010} is motivated not just by the persistent nature of the red-blue asymmetries in the single Gaussian fit but, more importantly, by the fact that these single Gaussian R-B asymmetries (which can be thought of as a proxy for the goodness-of-fit $\chi^2$) are not uniform but show (physical) structuring (as can for example be seen in Figure 1(e) of \citealt{DePontieu2010}). Such physical structuring in a goodness-of-fit measure indicates that essential physics is lacking in the (single Gaussian fit) model and hence that a more complex model, such as the double Gaussian fit used by these authors, is needed. The values of the R-B asymmetries in the single Gaussian fit are used as the initial guess for the secondary component of the double Gaussian fit which seems an appropriate initial guess. However, it is worth noting here that fixing the center position of the secondary component relatively far out in the blue wing in this way almost automatically results in (small) period Doppler shifts on top of a large, persistent blue shift, or, in other words, the periodic flow model.

We also point out that the relatively small size of our numerical domain is likely to affect our results as well. In our simple model, the line profiles are dominated by the flow or wave components rather than the (surrounding) plasma at rest which potentially makes them more sensitive to the initial guess. In reality, it is likely the reverse, with the plasma at rest along the LOS dominating the emission rather than the perturbed plasma (flow or wave) which might lead to a more stable fitting of (at least) the primary component.

Finally, we emphasise how different some of the observational signatures can look from the actual underlying theoretical model. This was nicely illustrated by the apparent `shocks' in the Fe~{\sc{xii}} 193 {\AA} lines along one particular LOS, when the velocity perturbations in our simulations are relatively small and show no signs of shocks. In this paper, we only varied one of the parameters of the multistrand model namely the phase difference between the oscillations in neighbouring strands. Even this small change leads to noticeable changes in the observational signatures. In reality there would of course be far more variations possible; the width of the strands, the number of the strands, the amplitude and periods of the perturbations could all be varied, as well as the background temperature and density of the strands.

\section{Conclusions}\label{sec:concl}

As was already apparent from the inconclusive observational debate on the nature of the observed propagating disturbances (PCDs), finding unique and robust observational signatures to distinguish the propagating, slow magneto acoustic wave model and the quasi-periodic upflow interpretation is highly non-trivial. Even the basic model presented here clearly shows that observational signatures are highly model-dependent and distinguishing between slow, propagating waves and periodic flows might simply not be possible. 

Possible observational signatures which might allow distinguishing the periodic upflow and slow propagating wave models are:
\begin{itemize}
\item The average line widths for the wave model vary rapidly as a function of LOS angle for shallow LOS angles;
\item The ratio of the line width amplitudes to the Doppler velocity amplitudes ($\Delta w / \Delta v_{{\rm Dopp}}$) for flows is relatively large, especially in the Fe~{\sc{ix}} line;
\item The ratio of the mean to the amplitude of the Doppler perturbations is larger than one for the flow model and in our upflow model was the same in the Fe~{\sc{ix}} and Fe~{\sc{xii}} spectral lines.
\end{itemize}
This last property appears to be the most robust signature. Linked to this is the absence of frequency-doubling in the observed line-widths which (as suggested by \opencite{Verwichte2010}) could indicate the presence of a steady upflow component along the LOS and hence, in this interpretation, would make a flow model more appropriate.  We do caution however, that the `switch' ({\it i.e.}~the critical mean/amplitude ratio) between the flow and wave model suggested here might be model dependent and hence a more comprehensive parameter-space investigation would be useful to confirm this. We also remind the reader that the results obtained in this study are based on single Gaussian fits to the spectral lines.

The most likely scenario able to account for the observational results (and discrepancies) appears to be the ``dual'' model, where upflows at the very base of the coronal loops generate a slow magneto acoustic wave which travels along the coronal loop. Such a dual model accommodates the differences between the spectral observations (which are often situated near the loop footpoint and find slightly lower speeds - {\it i.e.}~they are mainly seeing the flow component of the dual model) and the (coronal)  imaging observations (observing the more extended coronal structures showing PCDs travelling at constants speeds of the order of the local sound speed, {\it i.e.}~the wave component of the dual model).

\begin{acknowledgements}
IDM acknowledges support of a Royal Society University Research Fellowship and a KU~Leuven Research Council senior research fellowship (SF/12/008).The research leading to these results has also received funding from the European Commissionâ Seventh Framework Programme (FP7/ 2007-2013) under the grant agreement SOLSPANET (project No. 269299,  \url{www.solspanet.eu/solspanet}). TVD has been sponsored by an Odysseus grant of the FWO Vlaanderen. The research was performed in the context of the IAP P7/08 CHARM (Belspo)
and the GOA-2015-014 (KU~Leuven). TVD acknowledges the funding from the
FP7 ERG grant with number 276808.The authors would like to thank Dr S.W.~McIntosh for helpful discussions.
\end{acknowledgements}

\bibliographystyle{spr-mp-sola}
\bibliography{wavesflows}          

\newcommand{\noop}[1]{}
\begin{thebibliography}{49}
\ifx\bisbn     \undefined \def\bisbn  #1{ISBN #1}\fi
\ifx\binits    \undefined \def\binits#1{#1}\fi
\ifx\bauthor   \undefined \def\bauthor#1{#1}\fi
\ifx\batitle   \undefined \def\batitle#1{#1}\fi
\ifx\bjtitle   \undefined \def\bjtitle#1{\textit{#1}}\fi
\ifx\bvolume   \undefined \def\bvolume#1{\textbf{#1}}\fi
\ifx\byear     \undefined \def\byear#1{#1}\fi
\ifx\bissue    \undefined \def\bissue#1{#1}\fi
\ifx\bfpage    \undefined \def\bfpage#1{#1}\fi
\ifx\blpage    \undefined \def\blpage #1{#1}\fi
\ifx\burl      \undefined \def\burl#1{\textsf{#1}}\fi
\ifx\href      \undefined \def\href#1#2{\textsf{#2}}\fi
\ifx\betal     \undefined \def\betal{\textit{et al.}}\fi
\ifx\bctitle   \undefined \def\bctitle#1{#1}\fi
\ifx\beditor   \undefined \def\beditor#1{#1}\fi
\ifx\bbtitle   \undefined \def\bbtitle#1{\textit{#1}}\fi
\ifx\bedition  \undefined \def\bedition#1{#1}\fi
\ifx\bseriesno \undefined \def\bseriesno#1{\textbf{#1}}\fi
\ifx\blocation \undefined \def\blocation#1{#1}\fi
\ifx\bsertitle \undefined \def\bsertitle#1{\textit{#1}}\fi
\ifx\bsnm      \undefined \def\bsnm#1{#1}\fi
\ifx\bsuffix   \undefined \def\bsuffix#1{#1}\fi
\ifx\bparticle \undefined \def\bparticle#1{#1}\fi
\ifx\barticle  \undefined \def\barticle#1{}\fi
\ifx\binstitute  \undefined \def\binstitute#1{#1}\fi
\ifx\bpublisher  \undefined \def\bpublisher#1{#1}\fi
\ifx\doiurl    \undefined
  \def\doiurl#1{\href{http://dx.doi.org/#1}{\textsf{DOI}}}\fi
\ifx\arxivurl  \undefined
  \def\arxivurl#1{\href{http://arxiv.org/abs/#1}{\textsf{arXiv}}}\fi
\ifx\adsurl    \undefined
  \def\adsurl#1{\href{http://adsabs.harvard.edu/abs/#1}{\textsf{ADS}}}\fi
\ifx\botherref \undefined \def\botherref#1{}\fi
\ifx\url       \undefined \def\url#1{\textsf{#1}}\fi
\ifx\bchapter  \undefined \def\bchapter#1{}\fi
\ifx\bbook     \undefined \def\bbook#1{}\fi
\ifx\bcomment  \undefined \def\bcomment#1{#1}\fi
\ifx\oauthor   \undefined \def\oauthor#1{#1}\fi
\ifx\citeauthoryear \undefined\def \citeauthoryear#1{#1}\fi
\def\endbibitem {}
\ifx\bconflocation  \undefined \def\bconflocation#1{#1} \fi

\bibitem[\protect\citeauthoryear{{Antolin} and {Van
  Doorsselaere}}{2013}]{Antolin2013}
\begin{barticle}
\bauthor{\bsnm{{Antolin}}, \binits{P.}},
\bauthor{\bsnm{{Van Doorsselaere}}, \binits{T.}}:
\byear{2013},
\batitle{{Line-of-sight geometrical and instrumental resolution effects on
  intensity perturbations by sausage modes}}.
\bjtitle{\aap}
\bvolume{555},
\bfpage{A74}.
\doiurl{10.1051/0004-6361/201220784}.
\adsurl{http://cdsads.u-strasbg.fr/abs/2013A\%26A...555A..74A}.
\end{barticle}
\endbibitem

\bibitem[\protect\citeauthoryear{{Arber} \textit{et~al.}}{2001}]{Arber2001}
\begin{barticle}
\bauthor{\bsnm{{Arber}}, \binits{T.D.}},
\bauthor{\bsnm{{Longbottom}}, \binits{A.W.}},
\bauthor{\bsnm{{Gerrard}}, \binits{C.L.}},
\bauthor{\bsnm{{Milne}}, \binits{A.M.}}:
\byear{2001},
\batitle{{A Staggered Grid, Lagrangian-Eulerian Remap Code for 3-D MHD
  Simulations}}.
\bjtitle{J. Comput. Phys.}
\bvolume{171},
\bfpage{151}.
\doiurl{10.1006/jcph.2001.6780}.
\adsurl{2001JCoPh.171..151A}.
\end{barticle}
\endbibitem

\bibitem[\protect\citeauthoryear{{Banerjee}, {Gupta}, and
  {Teriaca}}{2011}]{Banerjee2011}
\begin{barticle}
\bauthor{\bsnm{{Banerjee}}, \binits{D.}},
\bauthor{\bsnm{{Gupta}}, \binits{G.R.}},
\bauthor{\bsnm{{Teriaca}}, \binits{L.}}:
\byear{2011},
\batitle{{Propagating MHD Waves in Coronal Holes}}.
\bjtitle{Space Sci. Rev.}
\bvolume{158},
\bfpage{267}.
\doiurl{10.1007/s11214-010-9698-z}.
\adsurl{2011SSRv..158..267B}.
\end{barticle}
\endbibitem

\bibitem[\protect\citeauthoryear{{Banerjee}, {O'Shea}, and
  {Doyle}}{2000}]{Banerjee2000}
\begin{barticle}
\bauthor{\bsnm{{Banerjee}}, \binits{D.}},
\bauthor{\bsnm{{O'Shea}}, \binits{E.}},
\bauthor{\bsnm{{Doyle}}, \binits{J.G.}}:
\byear{2000},
\batitle{{Long-Period Oscillations in Polar Plumes as Observed by cds on
  Soho}}.
\bjtitle{Solar Phys.}
\bvolume{196},
\bfpage{63}.
\adsurl{2000SoPh..196...63B}.
\end{barticle}
\endbibitem

\bibitem[\protect\citeauthoryear{{Berghmans} and
  {Clette}}{1999}]{Berghmans1999}
\begin{barticle}
\bauthor{\bsnm{{Berghmans}}, \binits{D.}},
\bauthor{\bsnm{{Clette}}, \binits{F.}}:
\byear{1999},
\batitle{{Active region EUV transient brightenings - First Results by EIT of
  SOHO JOP80}}.
\bjtitle{Solar Phys.}
\bvolume{186},
\bfpage{207}.
\adsurl{1999SoPh..186..207B}.
\end{barticle}
\endbibitem

\bibitem[\protect\citeauthoryear{{Bryans}, {Young}, and
  {Doschek}}{2010}]{Bryans2010}
\begin{barticle}
\bauthor{\bsnm{{Bryans}}, \binits{P.}},
\bauthor{\bsnm{{Young}}, \binits{P.R.}},
\bauthor{\bsnm{{Doschek}}, \binits{G.A.}}:
\byear{2010},
\batitle{{Multiple Component Outflows in an Active Region Observed with the EUV
  Imaging Spectrometer on Hinode}}.
\bjtitle{Astrophys.J.}
\bvolume{715},
\bfpage{1012}.
\doiurl{10.1088/0004-637X/715/2/1012}.
\adsurl{2010ApJ...715.1012B}.
\end{barticle}
\endbibitem

\bibitem[\protect\citeauthoryear{{Culhane} \textit{et~al.}}{2007}]{Culhane2007}
\begin{barticle}
\bauthor{\bsnm{{Culhane}}, \binits{J.L.}},
\bauthor{\bsnm{{Harra}}, \binits{L.K.}},
\bauthor{\bsnm{{James}}, \binits{A.M.}},
\bauthor{\bsnm{{Al-Janabi}}, \binits{K.}},
\bauthor{\bsnm{{Bradley}}, \binits{L.J.}},
\bauthor{\bsnm{{Chaudry}}, \binits{R.A.}},
\bauthor{\bsnm{{Rees}}, \binits{K.}},
\bauthor{\bsnm{{Tandy}}, \binits{J.A.}},
\bauthor{\bsnm{{Thomas}}, \binits{P.}},
\bauthor{\bsnm{{Whillock}}, \binits{M.C.R.}},
\bauthor{\bsnm{{Winter}}, \binits{B.}},
\bauthor{\bsnm{{Doschek}}, \binits{G.A.}},
\bauthor{\bsnm{{Korendyke}}, \binits{C.M.}},
\bauthor{\bsnm{{Brown}}, \binits{C.M.}},
\bauthor{\bsnm{{Myers}}, \binits{S.}},
\bauthor{\bsnm{{Mariska}}, \binits{J.}},
\bauthor{\bsnm{{Seely}}, \binits{J.}},
\bauthor{\bsnm{{Lang}}, \binits{J.}},
\bauthor{\bsnm{{Kent}}, \binits{B.J.}},
\bauthor{\bsnm{{Shaughnessy}}, \binits{B.M.}},
\bauthor{\bsnm{{Young}}, \binits{P.R.}},
\bauthor{\bsnm{{Simnett}}, \binits{G.M.}},
\bauthor{\bsnm{{Castelli}}, \binits{C.M.}},
\bauthor{\bsnm{{Mahmoud}}, \binits{S.}},
\bauthor{\bsnm{{Mapson-Menard}}, \binits{H.}},
\bauthor{\bsnm{{Probyn}}, \binits{B.J.}},
\bauthor{\bsnm{{Thomas}}, \binits{R.J.}},
\bauthor{\bsnm{{Davila}}, \binits{J.}},
\bauthor{\bsnm{{Dere}}, \binits{K.}},
\bauthor{\bsnm{{Windt}}, \binits{D.}},
\bauthor{\bsnm{{Shea}}, \binits{J.}},
\bauthor{\bsnm{{Hagood}}, \binits{R.}},
\bauthor{\bsnm{{Moye}}, \binits{R.}},
\bauthor{\bsnm{{Hara}}, \binits{H.}},
\bauthor{\bsnm{{Watanabe}}, \binits{T.}},
\bauthor{\bsnm{{Matsuzaki}}, \binits{K.}},
\bauthor{\bsnm{{Kosugi}}, \binits{T.}},
\bauthor{\bsnm{{Hansteen}}, \binits{V.}},
\bauthor{\bsnm{{Wikstol}}, \binits{{\O}.}}:
\byear{2007},
\batitle{{The EUV Imaging Spectrometer for Hinode}}.
\bjtitle{\solphys}
\bvolume{243},
\bfpage{19}.
\doiurl{10.1007/s01007-007-0293-1}.
\adsurl{http://cdsads.u-strasbg.fr/abs/2007SoPh..243...19C}.
\end{barticle}
\endbibitem

\bibitem[\protect\citeauthoryear{{De Moortel}}{2009}]{IDM09}
\begin{barticle}
\bauthor{\bsnm{{De Moortel}}, \binits{I.}}:
\byear{2009},
\batitle{{Longitudinal Waves in Coronal Loops}}.
\bjtitle{Space Sci. Rev.}
\bvolume{149},
\bfpage{65}.
\doiurl{10.1007/s11214-009-9526-5}.
\adsurl{2009SSRv..149...65D}.
\end{barticle}
\endbibitem

\bibitem[\protect\citeauthoryear{{De Moortel} and {Bradshaw}}{2008}]{IDM08}
\begin{barticle}
\bauthor{\bsnm{{De Moortel}}, \binits{I.}},
\bauthor{\bsnm{{Bradshaw}}, \binits{S.J.}}:
\byear{2008},
\batitle{{Forward Modelling of Coronal Intensity Perturbations}}.
\bjtitle{\solphys}
\bvolume{252},
\bfpage{101}.
\doiurl{10.1007/s11207-008-9238-0}.
\adsurl{http://cdsads.u-strasbg.fr/abs/2008SoPh..252..101D}.
\end{barticle}
\endbibitem

\bibitem[\protect\citeauthoryear{{De Moortel} and {Hood}}{2003}]{IDM03}
\begin{barticle}
\bauthor{\bsnm{{De Moortel}}, \binits{I.}},
\bauthor{\bsnm{{Hood}}, \binits{A.W.}}:
\byear{2003},
\batitle{{The damping of slow MHD waves in solar coronal magnetic fields}}.
\bjtitle{Astron. Astrophys.}
\bvolume{408},
\bfpage{755}.
\doiurl{10.1051/0004-6361:20030984}.
\adsurl{2003A\%26A...408..755D}.
\end{barticle}
\endbibitem

\bibitem[\protect\citeauthoryear{{De Moortel} and {Hood}}{2004}]{IDM04}
\begin{barticle}
\bauthor{\bsnm{{De Moortel}}, \binits{I.}},
\bauthor{\bsnm{{Hood}}, \binits{A.W.}}:
\byear{2004},
\batitle{{The damping of slow MHD waves in solar coronal magnetic fields. II.
  The effect of gravitational stratification and field line divergence}}.
\bjtitle{Astron. Astrophys.}
\bvolume{415},
\bfpage{705}.
\doiurl{10.1051/0004-6361:20034233}.
\adsurl{2004A\%26A...415..705D}.
\end{barticle}
\endbibitem

\bibitem[\protect\citeauthoryear{{De Moortel} and
  {Nakariakov}}{2012}]{review:DeMoortelNakariakov2012}
\begin{barticle}
\bauthor{\bsnm{{De Moortel}}, \binits{I.}},
\bauthor{\bsnm{{Nakariakov}}, \binits{V.M.}}:
\byear{2012},
\batitle{{Magnetohydrodynamic waves and coronal seismology: an overview of
  recent results}}.
\bjtitle{Phil. Trans. Roy. Soc. London A}
\bvolume{370},
\bfpage{3193}.
\doiurl{10.1098/rsta.2011.0640}.
\adsurl{2012RSPTA.370.3193D}.
\end{barticle}
\endbibitem

\bibitem[\protect\citeauthoryear{{De Moortel}, {Ireland}, and
  {Walsh}}{2000}]{IDM00}
\begin{barticle}
\bauthor{\bsnm{{De Moortel}}, \binits{I.}},
\bauthor{\bsnm{{Ireland}}, \binits{J.}},
\bauthor{\bsnm{{Walsh}}, \binits{R.W.}}:
\byear{2000},
\batitle{{Observation of oscillations in coronal loops}}.
\bjtitle{Astron. Astrophys.}
\bvolume{355},
\bfpage{L23}.
\adsurl{2000A\%26A...355L..23D}.
\end{barticle}
\endbibitem

\bibitem[\protect\citeauthoryear{{De Moortel}
  \textit{et~al.}}{2002a}]{IDM2002b}
\begin{barticle}
\bauthor{\bsnm{{De Moortel}}, \binits{I.}},
\bauthor{\bsnm{{Ireland}}, \binits{J.}},
\bauthor{\bsnm{{Walsh}}, \binits{R.W.}},
\bauthor{\bsnm{{Hood}}, \binits{A.W.}}:
\byear{2002}a,
\batitle{{Longitudinal intensity oscillations in coronal loops observed with
  TRACE I. Overview of Measured Parameters}}.
\bjtitle{\solphys}
\bvolume{209},
\bfpage{61}.
\doiurl{10.1023/A:1020956421063}.
\adsurl{http://cdsads.u-strasbg.fr/abs/2002SoPh..209...61D}.
\end{barticle}
\endbibitem

\bibitem[\protect\citeauthoryear{{De Moortel}
  \textit{et~al.}}{2002b}]{IDM2002a}
\begin{barticle}
\bauthor{\bsnm{{De Moortel}}, \binits{I.}},
\bauthor{\bsnm{{Hood}}, \binits{A.W.}},
\bauthor{\bsnm{{Ireland}}, \binits{J.}},
\bauthor{\bsnm{{Walsh}}, \binits{R.W.}}:
\byear{2002}b,
\batitle{{Longitudinal intensity oscillations in coronal loops observed with
  TRACE II. Discussion of Measured Parameters}}.
\bjtitle{\solphys}
\bvolume{209},
\bfpage{89}.
\doiurl{10.1023/A:1020960505133}.
\adsurl{http://cdsads.u-strasbg.fr/abs/2002SoPh..209...89D}.
\end{barticle}
\endbibitem

\bibitem[\protect\citeauthoryear{{De Pontieu} and
  {McIntosh}}{2010}]{DePontieu2010}
\begin{barticle}
\bauthor{\bsnm{{De Pontieu}}, \binits{B.}},
\bauthor{\bsnm{{McIntosh}}, \binits{S.W.}}:
\byear{2010},
\batitle{{Quasi-periodic Propagating Signals in the Solar Corona: The Signature
  of Magnetoacoustic Waves or High-velocity Upflows?}}
\bjtitle{Astrophys. J.}
\bvolume{722},
\bfpage{1013}.
\doiurl{10.1088/0004-637X/722/2/1013}.
\adsurl{2010ApJ...722.1013D}.
\end{barticle}
\endbibitem

\bibitem[\protect\citeauthoryear{{De Pontieu}
  \textit{et~al.}}{2009}]{DePontieu2009}
\begin{barticle}
\bauthor{\bsnm{{De Pontieu}}, \binits{B.}},
\bauthor{\bsnm{{McIntosh}}, \binits{S.W.}},
\bauthor{\bsnm{{Hansteen}}, \binits{V.H.}},
\bauthor{\bsnm{{Schrijver}}, \binits{C.J.}}:
\byear{2009},
\batitle{{Observing the Roots of Solar Coronal Heating in the Chromosphere}}.
\bjtitle{Astrophys. J. Lett.}
\bvolume{701},
\bfpage{L1}.
\doiurl{10.1088/0004-637X/701/1/L1}.
\adsurl{2009ApJ...701L...1D}.
\end{barticle}
\endbibitem

\bibitem[\protect\citeauthoryear{{De Pontieu}
  \textit{et~al.}}{2011}]{DePontieu2011}
\begin{barticle}
\bauthor{\bsnm{{De Pontieu}}, \binits{B.}},
\bauthor{\bsnm{{McIntosh}}, \binits{S.W.}},
\bauthor{\bsnm{{Carlsson}}, \binits{M.}},
\bauthor{\bsnm{{Hansteen}}, \binits{V.H.}},
\bauthor{\bsnm{{Tarbell}}, \binits{T.D.}},
\bauthor{\bsnm{{Boerner}}, \binits{P.}},
\bauthor{\bsnm{{Martinez-Sykora}}, \binits{J.}},
\bauthor{\bsnm{{Schrijver}}, \binits{C.J.}},
\bauthor{\bsnm{{Title}}, \binits{A.M.}}:
\byear{2011},
\batitle{{The Origins of Hot Plasma in the Solar Corona}}.
\bjtitle{Science}
\bvolume{331}.
\doiurl{10.1126/science.1197738}.
\adsurl{2011Sci...331...55D}.
\end{barticle}
\endbibitem

\bibitem[\protect\citeauthoryear{{DeForest} and {Gurman}}{1998}]{DeForest1998}
\begin{barticle}
\bauthor{\bsnm{{DeForest}}, \binits{C.E.}},
\bauthor{\bsnm{{Gurman}}, \binits{J.B.}}:
\byear{1998},
\batitle{{Observation of Quasi-periodic Compressive Waves in Solar Polar
  Plumes}}.
\bjtitle{\apjl}
\bvolume{501},
\bfpage{L217}.
\doiurl{10.1086/311460}.
\adsurl{http://cdsads.u-strasbg.fr/abs/1998ApJ...501L.217D}.
\end{barticle}
\endbibitem

\bibitem[\protect\citeauthoryear{{Del Zanna}}{2008}]{DelZanna2008}
\begin{barticle}
\bauthor{\bsnm{{Del Zanna}}, \binits{G.}}:
\byear{2008},
\batitle{{Flows in active region loops observed by Hinode EIS}}.
\bjtitle{Astron. Astrophys.}
\bvolume{481},
\bfpage{L49}.
\doiurl{10.1051/0004-6361:20079087}.
\adsurl{2008A\%26A...481L..49D}.
\end{barticle}
\endbibitem

\bibitem[\protect\citeauthoryear{{Doschek} \textit{et~al.}}{2008}]{Doschek2008}
\begin{barticle}
\bauthor{\bsnm{{Doschek}}, \binits{G.A.}},
\bauthor{\bsnm{{Warren}}, \binits{H.P.}},
\bauthor{\bsnm{{Mariska}}, \binits{J.T.}},
\bauthor{\bsnm{{Muglach}}, \binits{K.}},
\bauthor{\bsnm{{Culhane}}, \binits{J.L.}},
\bauthor{\bsnm{{Hara}}, \binits{H.}},
\bauthor{\bsnm{{Watanabe}}, \binits{T.}}:
\byear{2008},
\batitle{{Flows and Nonthermal Velocities in Solar Active Regions Observed with
  the EUV Imaging Spectrometer on Hinode: A Tracer of Active Region Sources of
  Heliospheric Magnetic Fields?}}
\bjtitle{Astrophys.J.}
\bvolume{686},
\bfpage{1362}.
\doiurl{10.1086/591724}.
\adsurl{2008ApJ...686.1362D}.
\end{barticle}
\endbibitem

\bibitem[\protect\citeauthoryear{{Hara} \textit{et~al.}}{2008}]{Hara2008}
\begin{barticle}
\bauthor{\bsnm{{Hara}}, \binits{H.}},
\bauthor{\bsnm{{Watanabe}}, \binits{T.}},
\bauthor{\bsnm{{Harra}}, \binits{L.K.}},
\bauthor{\bsnm{{Culhane}}, \binits{J.L.}},
\bauthor{\bsnm{{Young}}, \binits{P.R.}},
\bauthor{\bsnm{{Mariska}}, \binits{J.T.}},
\bauthor{\bsnm{{Doschek}}, \binits{G.A.}}:
\byear{2008},
\batitle{{Coronal Plasma Motions near Footpoints of Active Region Loops
  Revealed from Spectroscopic Observations with Hinode EIS}}.
\bjtitle{Astrophys.J.}
\bvolume{678},
\bfpage{L67}.
\doiurl{10.1086/588252}.
\adsurl{2008ApJ...678L..67H}.
\end{barticle}
\endbibitem

\bibitem[\protect\citeauthoryear{{Harra} \textit{et~al.}}{2008}]{Harra2008}
\begin{barticle}
\bauthor{\bsnm{{Harra}}, \binits{L.K.}},
\bauthor{\bsnm{{Sakao}}, \binits{T.}},
\bauthor{\bsnm{{Mandrini}}, \binits{C.H.}},
\bauthor{\bsnm{{Hara}}, \binits{H.}},
\bauthor{\bsnm{{Imada}}, \binits{S.}},
\bauthor{\bsnm{{Young}}, \binits{P.R.}},
\bauthor{\bsnm{{van Driel-Gesztelyi}}, \binits{L.}},
\bauthor{\bsnm{{Baker}}, \binits{D.}}:
\byear{2008},
\batitle{{Outflows at the Edges of Active Regions: Contribution to Solar Wind
  Formation?}}
\bjtitle{Astrophys.J.}
\bvolume{676},
\bfpage{L147}.
\doiurl{10.1086/587485}.
\adsurl{2008ApJ...676L.147H}.
\end{barticle}
\endbibitem

\bibitem[\protect\citeauthoryear{{He} \textit{et~al.}}{2010}]{He2010}
\begin{barticle}
\bauthor{\bsnm{{He}}, \binits{J.-S.}},
\bauthor{\bsnm{{Marsch}}, \binits{E.}},
\bauthor{\bsnm{{Tu}}, \binits{C.-Y.}},
\bauthor{\bsnm{{Guo}}, \binits{L.-J.}},
\bauthor{\bsnm{{Tian}}, \binits{H.}}:
\byear{2010},
\batitle{{Intermittent outflows at the edge of an active region - a possible
  source of the solar wind?}}
\bjtitle{Astron. Astrophys.}
\bvolume{516},
\bfpage{A14}.
\doiurl{10.1051/0004-6361/200913712}.
\adsurl{2010A\%26A...516A..14H}.
\end{barticle}
\endbibitem

\bibitem[\protect\citeauthoryear{{Lemen} \textit{et~al.}}{2012}]{Lemen2012}
\begin{barticle}
\bauthor{\bsnm{{Lemen}}, \binits{J.R.}},
\bauthor{\bsnm{{Title}}, \binits{A.M.}},
\bauthor{\bsnm{{Akin}}, \binits{D.J.}},
\bauthor{\bsnm{{Boerner}}, \binits{P.F.}},
\bauthor{\bsnm{{Chou}}, \binits{C.}},
\bauthor{\bsnm{{Drake}}, \binits{J.F.}},
\bauthor{\bsnm{{Duncan}}, \binits{D.W.}},
\bauthor{\bsnm{{Edwards}}, \binits{C.G.}},
\bauthor{\bsnm{{Friedlaender}}, \binits{F.M.}},
\bauthor{\bsnm{{Heyman}}, \binits{G.F.}},
\bauthor{\bsnm{{Hurlburt}}, \binits{N.E.}},
\bauthor{\bsnm{{Katz}}, \binits{N.L.}},
\bauthor{\bsnm{{Kushner}}, \binits{G.D.}},
\bauthor{\bsnm{{Levay}}, \binits{M.}},
\bauthor{\bsnm{{Lindgren}}, \binits{R.W.}},
\bauthor{\bsnm{{Mathur}}, \binits{D.P.}},
\bauthor{\bsnm{{McFeaters}}, \binits{E.L.}},
\bauthor{\bsnm{{Mitchell}}, \binits{S.}},
\bauthor{\bsnm{{Rehse}}, \binits{R.A.}},
\bauthor{\bsnm{{Schrijver}}, \binits{C.J.}},
\bauthor{\bsnm{{Springer}}, \binits{L.A.}},
\bauthor{\bsnm{{Stern}}, \binits{R.A.}},
\bauthor{\bsnm{{Tarbell}}, \binits{T.D.}},
\bauthor{\bsnm{{Wuelser}}, \binits{J.-P.}},
\bauthor{\bsnm{{Wolfson}}, \binits{C.J.}},
\bauthor{\bsnm{{Yanari}}, \binits{C.}},
\bauthor{\bsnm{{Bookbinder}}, \binits{J.A.}},
\bauthor{\bsnm{{Cheimets}}, \binits{P.N.}},
\bauthor{\bsnm{{Caldwell}}, \binits{D.}},
\bauthor{\bsnm{{Deluca}}, \binits{E.E.}},
\bauthor{\bsnm{{Gates}}, \binits{R.}},
\bauthor{\bsnm{{Golub}}, \binits{L.}},
\bauthor{\bsnm{{Park}}, \binits{S.}},
\bauthor{\bsnm{{Podgorski}}, \binits{W.A.}},
\bauthor{\bsnm{{Bush}}, \binits{R.I.}},
\bauthor{\bsnm{{Scherrer}}, \binits{P.H.}},
\bauthor{\bsnm{{Gummin}}, \binits{M.A.}},
\bauthor{\bsnm{{Smith}}, \binits{P.}},
\bauthor{\bsnm{{Auker}}, \binits{G.}},
\bauthor{\bsnm{{Jerram}}, \binits{P.}},
\bauthor{\bsnm{{Pool}}, \binits{P.}},
\bauthor{\bsnm{{Soufli}}, \binits{R.}},
\bauthor{\bsnm{{Windt}}, \binits{D.L.}},
\bauthor{\bsnm{{Beardsley}}, \binits{S.}},
\bauthor{\bsnm{{Clapp}}, \binits{M.}},
\bauthor{\bsnm{{Lang}}, \binits{J.}},
\bauthor{\bsnm{{Waltham}}, \binits{N.}}:
\byear{2012},
\batitle{{The Atmospheric Imaging Assembly (AIA) on the Solar Dynamics
  Observatory (SDO)}}.
\bjtitle{\solphys}
\bvolume{275},
\bfpage{17}.
\doiurl{10.1007/s11207-011-9776-8}.
\adsurl{2012SoPh..275...17L}.
\end{barticle}
\endbibitem

\bibitem[\protect\citeauthoryear{{Marsh} and {Walsh}}{2009}]{Marsh2009b}
\begin{barticle}
\bauthor{\bsnm{{Marsh}}, \binits{M.S.}},
\bauthor{\bsnm{{Walsh}}, \binits{R.W.}}:
\byear{2009},
\batitle{{Using HINODE/Extreme-Ultraviolet Imaging Spectrometer to Confirm a
  Seismologically Inferred Coronal Temperature}}.
\bjtitle{\apjl}
\bvolume{706},
\bfpage{L76}.
\doiurl{10.1088/0004-637X/706/1/L76}.
\adsurl{http://cdsads.u-strasbg.fr/abs/2009ApJ...706L..76M}.
\end{barticle}
\endbibitem

\bibitem[\protect\citeauthoryear{{Marsh}, {Walsh}, and
  {Plunkett}}{2009}]{Marsh2009a}
\begin{barticle}
\bauthor{\bsnm{{Marsh}}, \binits{M.S.}},
\bauthor{\bsnm{{Walsh}}, \binits{R.W.}},
\bauthor{\bsnm{{Plunkett}}, \binits{S.}}:
\byear{2009},
\batitle{{Three-dimensional Coronal Slow Modes: Toward Three-dimensional
  Seismology}}.
\bjtitle{\apj}
\bvolume{697},
\bfpage{1674}.
\doiurl{10.1088/0004-637X/697/2/1674}.
\adsurl{http://cdsads.u-strasbg.fr/abs/2009ApJ...697.1674M}.
\end{barticle}
\endbibitem

\bibitem[\protect\citeauthoryear{{McIntosh}}{2012}]{McIntosh2012REV}
\begin{barticle}
\bauthor{\bsnm{{McIntosh}}, \binits{S.W.}}:
\byear{2012},
\batitle{{Recent Observations of Plasma and Alfv{\'e}nic Wave Energy Injection
  at the Base of the Fast Solar Wind}}.
\bjtitle{\ssr}
\bvolume{172},
\bfpage{69}.
\doiurl{10.1007/s11214-012-9889-x}.
\adsurl{http://cdsads.u-strasbg.fr/abs/2012SSRv..172...69M}.
\end{barticle}
\endbibitem

\bibitem[\protect\citeauthoryear{{McIntosh} and {De
  Pontieu}}{2009a}]{McIntosh2009a}
\begin{barticle}
\bauthor{\bsnm{{McIntosh}}, \binits{S.W.}},
\bauthor{\bsnm{{De Pontieu}}, \binits{B.}}:
\byear{2009}a,
\batitle{{High-Speed Transition Region and Coronal Upflows in the Quiet Sun}}.
\bjtitle{Astrophys.J.}
\bvolume{707},
\bfpage{524}.
\doiurl{10.1088/0004-637X/707/1/524}.
\adsurl{2009ApJ...707..524M}.
\end{barticle}
\endbibitem

\bibitem[\protect\citeauthoryear{{McIntosh} and {De
  Pontieu}}{2009b}]{McIntosh2009b}
\begin{barticle}
\bauthor{\bsnm{{McIntosh}}, \binits{S.W.}},
\bauthor{\bsnm{{De Pontieu}}, \binits{B.}}:
\byear{2009}b,
\batitle{{Observing Episodic Coronal Heating Events Rooted in Chromospheric
  Activity}}.
\bjtitle{Astrophys.J.}
\bvolume{706},
\bfpage{L80}.
\doiurl{10.1088/0004-637X/706/1/L80}.
\adsurl{2009ApJ...706L..80M}.
\end{barticle}
\endbibitem

\bibitem[\protect\citeauthoryear{{McIntosh}
  \textit{et~al.}}{2010}]{McIntosh2010}
\begin{barticle}
\bauthor{\bsnm{{McIntosh}}, \binits{S.W.}},
\bauthor{\bsnm{{Innes}}, \binits{D.E.}},
\bauthor{\bsnm{{de Pontieu}}, \binits{B.}},
\bauthor{\bsnm{{Leamon}}, \binits{R.J.}}:
\byear{2010},
\batitle{{STEREO observations of quasi-periodically driven high velocity
  outflows in polar plumes}}.
\bjtitle{\aap}
\bvolume{510},
\bfpage{L2}.
\doiurl{10.1051/0004-6361/200913699}.
\adsurl{http://cdsads.u-strasbg.fr/abs/2010A\%26A...510L...2M}.
\end{barticle}
\endbibitem

\bibitem[\protect\citeauthoryear{{McIntosh}
  \textit{et~al.}}{2012}]{McIntosh2012}
\begin{barticle}
\bauthor{\bsnm{{McIntosh}}, \binits{S.W.}},
\bauthor{\bsnm{{Tian}}, \binits{H.}},
\bauthor{\bsnm{{Sechler}}, \binits{M.}},
\bauthor{\bsnm{{De Pontieu}}, \binits{B.}}:
\byear{2012},
\batitle{{On the Doppler Velocity of Emission Line Profiles Formed in the
  ''Coronal Contraflow'' that Is the Chromosphere-Corona Mass Cycle}}.
\bjtitle{\apj}
\bvolume{749},
\bfpage{60}.
\doiurl{10.1088/0004-637X/749/1/60}.
\adsurl{http://cdsads.u-strasbg.fr/abs/2012ApJ...749...60M}.
\end{barticle}
\endbibitem

\bibitem[\protect\citeauthoryear{{Nishizuka} and {Hara}}{2011}]{Nishizuka2011}
\begin{barticle}
\bauthor{\bsnm{{Nishizuka}}, \binits{N.}},
\bauthor{\bsnm{{Hara}}, \binits{H.}}:
\byear{2011},
\batitle{{Spectroscopic Observations of Continuous Outflows and Propagating
  Waves from NOAA 10942 with Extreme Ultraviolet Imaging Spectrometer/Hinode}}.
\bjtitle{\apjl}
\bvolume{737},
\bfpage{L43}.
\doiurl{10.1088/2041-8205/737/2/L43}.
\adsurl{http://cdsads.u-strasbg.fr/abs/2011ApJ...737L..43N}.
\end{barticle}
\endbibitem

\bibitem[\protect\citeauthoryear{{Ofman}, {Nakariakov}, and
  {Sehgal}}{2000}]{Ofman00}
\begin{barticle}
\bauthor{\bsnm{{Ofman}}, \binits{L.}},
\bauthor{\bsnm{{Nakariakov}}, \binits{V.M.}},
\bauthor{\bsnm{{Sehgal}}, \binits{N.}}:
\byear{2000},
\batitle{{Dissipation of Slow Magnetosonic Waves in Coronal Plumes}}.
\bjtitle{Astrophys.J.}
\bvolume{533},
\bfpage{1071}.
\doiurl{10.1086/308691}.
\adsurl{2000ApJ...533.1071O}.
\end{barticle}
\endbibitem

\bibitem[\protect\citeauthoryear{{Ofman}, {Wang}, and
  {Davila}}{2012}]{Ofman2012}
\begin{barticle}
\bauthor{\bsnm{{Ofman}}, \binits{L.}},
\bauthor{\bsnm{{Wang}}, \binits{T.J.}},
\bauthor{\bsnm{{Davila}}, \binits{J.M.}}:
\byear{2012},
\batitle{{Slow Magnetosonic Waves and Fast Flows in Active Region Loops}}.
\bjtitle{\apj}
\bvolume{754},
\bfpage{111}.
\doiurl{10.1088/0004-637X/754/2/111}.
\adsurl{http://cdsads.u-strasbg.fr/abs/2012ApJ...754..111O}.
\end{barticle}
\endbibitem

\bibitem[\protect\citeauthoryear{{Ofman} \textit{et~al.}}{1997}]{Ofman97}
\begin{barticle}
\bauthor{\bsnm{{Ofman}}, \binits{L.}},
\bauthor{\bsnm{{Romoli}}, \binits{M.}},
\bauthor{\bsnm{{Poletto}}, \binits{G.}},
\bauthor{\bsnm{{Noci}}, \binits{G.}},
\bauthor{\bsnm{{Kohl}}, \binits{J.L.}}:
\byear{1997},
\batitle{{Ultraviolet Coronagraph Spectrometer Observations of Density
  Fluctuations in the Solar Wind}}.
\bjtitle{Astrophys.J.}
\bvolume{491},
\bfpage{L111}.
\doiurl{10.1086/311067}.
\adsurl{1997ApJ...491L.111O}.
\end{barticle}
\endbibitem

\bibitem[\protect\citeauthoryear{{Owen}, {De Moortel}, and
  {Hood}}{2009}]{Owen2009}
\begin{barticle}
\bauthor{\bsnm{{Owen}}, \binits{N.R.}},
\bauthor{\bsnm{{De Moortel}}, \binits{I.}},
\bauthor{\bsnm{{Hood}}, \binits{A.W.}}:
\byear{2009},
\batitle{{Forward modelling to determine the observational signatures of
  propagating slow waves for TRACE, SoHO/CDS, and Hinode/EIS}}.
\bjtitle{\aap}
\bvolume{494},
\bfpage{339}.
\doiurl{10.1051/0004-6361:200810828}.
\adsurl{http://cdsads.u-strasbg.fr/abs/2009A\%26A...494..339O}.
\end{barticle}
\endbibitem

\bibitem[\protect\citeauthoryear{{Peter}}{2010}]{Peter2010}
\begin{barticle}
\bauthor{\bsnm{{Peter}}, \binits{H.}}:
\byear{2010},
\batitle{{Asymmetries of solar coronal extreme ultraviolet emission lines}}.
\bjtitle{Astron. Astrophys.}
\bvolume{521},
\bfpage{A51}.
\doiurl{10.1051/0004-6361/201014433}.
\adsurl{2010A\%26A...521A..51P}.
\end{barticle}
\endbibitem

\bibitem[\protect\citeauthoryear{{Sakao} \textit{et~al.}}{2007}]{Sakao2007}
\begin{barticle}
\bauthor{\bsnm{{Sakao}}, \binits{T.}},
\bauthor{\bsnm{{Kano}}, \binits{R.}},
\bauthor{\bsnm{{Narukage}}, \binits{N.}},
\bauthor{\bsnm{{Kotoku}}, \binits{J.}},
\bauthor{\bsnm{{Bando}}, \binits{T.}},
\bauthor{\bsnm{{DeLuca}}, \binits{E.E.}},
\bauthor{\bsnm{{Lundquist}}, \binits{L.L.}},
\bauthor{\bsnm{{Tsuneta}}, \binits{S.}},
\bauthor{\bsnm{{Harra}}, \binits{L.K.}},
\bauthor{\bsnm{{Katsukawa}}, \binits{Y.}},
\bauthor{\bsnm{{Kubo}}, \binits{M.}},
\bauthor{\bsnm{{Hara}}, \binits{H.}},
\bauthor{\bsnm{{Matsuzaki}}, \binits{K.}},
\bauthor{\bsnm{{Shimojo}}, \binits{M.}},
\bauthor{\bsnm{{Bookbinder}}, \binits{J.A.}},
\bauthor{\bsnm{{Golub}}, \binits{L.}},
\bauthor{\bsnm{{Korreck}}, \binits{K.E.}},
\bauthor{\bsnm{{Su}}, \binits{Y.}},
\bauthor{\bsnm{{Shibasaki}}, \binits{K.}},
\bauthor{\bsnm{{Shimizu}}, \binits{T.}},
\bauthor{\bsnm{{Nakatani}}, \binits{I.}}:
\byear{2007},
\batitle{{Continuous Plasma Outflows from the Edge of a Solar Active Region as
  a Possible Source of Solar Wind}}.
\bjtitle{Science}
\bvolume{318}.
\doiurl{10.1126/science.1147292}.
\adsurl{2007Sci...318.1585S}.
\end{barticle}
\endbibitem

\bibitem[\protect\citeauthoryear{{Schrijver}
  \textit{et~al.}}{1999}]{Schrijver99}
\begin{barticle}
\bauthor{\bsnm{{Schrijver}}, \binits{C.J.}},
\bauthor{\bsnm{{Title}}, \binits{A.M.}},
\bauthor{\bsnm{{Berger}}, \binits{T.E.}},
\bauthor{\bsnm{{Fletcher}}, \binits{L.}},
\bauthor{\bsnm{{Hurlburt}}, \binits{N.E.}},
\bauthor{\bsnm{{Nightingale}}, \binits{R.W.}},
\bauthor{\bsnm{{Shine}}, \binits{R.A.}},
\bauthor{\bsnm{{Tarbell}}, \binits{T.D.}},
\bauthor{\bsnm{{Wolfson}}, \binits{J.}},
\bauthor{\bsnm{{Golub}}, \binits{L.}},
\bauthor{\bsnm{{Bookbinder}}, \binits{J.A.}},
\bauthor{\bsnm{{Deluca}}, \binits{E.E.}},
\bauthor{\bsnm{{McMullen}}, \binits{R.A.}},
\bauthor{\bsnm{{Warren}}, \binits{H.P.}},
\bauthor{\bsnm{{Kankelborg}}, \binits{C.C.}},
\bauthor{\bsnm{{Handy}}, \binits{B.N.}},
\bauthor{\bsnm{{de Pontieu}}, \binits{B.}}:
\byear{1999},
\batitle{{A new view of the solar outer atmosphere by the Transition Region and
  Coronal Explorer}}.
\bjtitle{Solar Phys.}
\bvolume{187},
\bfpage{261}.
\doiurl{10.1023/A:1005194519642}.
\adsurl{1999SoPh..187..261S}.
\end{barticle}
\endbibitem

\bibitem[\protect\citeauthoryear{{Tian}, {McIntosh}, and {De
  Pontieu}}{2011}]{Tian2011}
\begin{barticle}
\bauthor{\bsnm{{Tian}}, \binits{H.}},
\bauthor{\bsnm{{McIntosh}}, \binits{S.W.}},
\bauthor{\bsnm{{De Pontieu}}, \binits{B.}}:
\byear{2011},
\batitle{{The Spectroscopic Signature of Quasi-periodic Upflows in Active
  Region Timeseries}}.
\bjtitle{Astrophys.J.}
\bvolume{727},
\bfpage{L37}.
\doiurl{10.1088/2041-8205/727/2/L37}.
\adsurl{2011ApJ...727L..37T}.
\end{barticle}
\endbibitem

\bibitem[\protect\citeauthoryear{{Ugarte-Urra} and
  {Warren}}{2011}]{Ugarte-Urra2011}
\begin{barticle}
\bauthor{\bsnm{{Ugarte-Urra}}, \binits{I.}},
\bauthor{\bsnm{{Warren}}, \binits{H.P.}}:
\byear{2011},
\batitle{{Temporal Variability of Active Region Outflows}}.
\bjtitle{Astrophys.J.}
\bvolume{730},
\bfpage{37}.
\doiurl{10.1088/0004-637X/730/1/37}.
\adsurl{2011ApJ...730...37U}.
\end{barticle}
\endbibitem

\bibitem[\protect\citeauthoryear{{Van Doorsselaere}
  \textit{et~al.}}{2011}]{VanDoorsselaere2011}
\begin{barticle}
\bauthor{\bsnm{{Van Doorsselaere}}, \binits{T.}},
\bauthor{\bsnm{{Wardle}}, \binits{N.}},
\bauthor{\bsnm{{Del Zanna}}, \binits{G.}},
\bauthor{\bsnm{{Jansari}}, \binits{K.}},
\bauthor{\bsnm{{Verwichte}}, \binits{E.}},
\bauthor{\bsnm{{Nakariakov}}, \binits{V.M.}}:
\byear{2011},
\batitle{{The First Measurement of the Adiabatic Index in the Solar Corona
  Using Time-dependent Spectroscopy of Hinode/EIS Observations}}.
\bjtitle{\apjl}
\bvolume{727},
\bfpage{L32}.
\doiurl{10.1088/2041-8205/727/2/L32}.
\adsurl{http://cdsads.u-strasbg.fr/abs/2011ApJ...727L..32V}.
\end{barticle}
\endbibitem

\bibitem[\protect\citeauthoryear{{Verwichte}
  \textit{et~al.}}{2010}]{Verwichte2010}
\begin{barticle}
\bauthor{\bsnm{{Verwichte}}, \binits{E.}},
\bauthor{\bsnm{{Marsh}}, \binits{M.}},
\bauthor{\bsnm{{Foullon}}, \binits{C.}},
\bauthor{\bsnm{{Van Doorsselaere}}, \binits{T.}},
\bauthor{\bsnm{{De Moortel}}, \binits{I.}},
\bauthor{\bsnm{{Hood}}, \binits{A.W.}},
\bauthor{\bsnm{{Nakariakov}}, \binits{V.M.}}:
\byear{2010},
\batitle{{Periodic Spectral Line Asymmetries in Solar Coronal Structures from
  Slow Magnetoacoustic Waves}}.
\bjtitle{\apjl}
\bvolume{724},
\bfpage{L194}.
\doiurl{10.1088/2041-8205/724/2/L194}.
\adsurl{2010ApJ...724L.194V}.
\end{barticle}
\endbibitem

\bibitem[\protect\citeauthoryear{{Wang}, {Ofman}, and
  {Davila}}{2012}]{Wang2012}
\begin{bchapter}
\bauthor{\bsnm{{Wang}}, \binits{T.}},
\bauthor{\bsnm{{Ofman}}, \binits{L.}},
\bauthor{\bsnm{{Davila}}, \binits{J.M.}}:
\byear{2012},
\bctitle{{Spectroscopic Diagnosis of Propagating Disturbances in Coronal Loops:
  Waves or flows?}}
In: \beditor{\bsnm{{Golub}}, \binits{L.}},
\beditor{\bsnm{{De Moortel}}, \binits{I.}},
\beditor{\bsnm{{Shimizu}}, \binits{T.}} (eds.)
\bbtitle{Fifth Hinode Science Meeting},
\bsertitle{Astronomical Society of the Pacific Conference Series}
\bseriesno{456},
\bfpage{91}.
\adsurl{http://cdsads.u-strasbg.fr/abs/2012ASPC..456...91W}.
\end{bchapter}
\endbibitem

\bibitem[\protect\citeauthoryear{{Wang}, {Ofman}, and
  {Davila}}{2013}]{Wang2013}
\begin{barticle}
\bauthor{\bsnm{{Wang}}, \binits{T.}},
\bauthor{\bsnm{{Ofman}}, \binits{L.}},
\bauthor{\bsnm{{Davila}}, \binits{J.M.}}:
\byear{2013},
\batitle{{Three-dimensional Magnetohydrodynamic Modeling of Propagating
  Disturbances in Fan-like Coronal Loops}}.
\bjtitle{\apjl}
\bvolume{775},
\bfpage{L23}.
\doiurl{10.1088/2041-8205/775/1/L23}.
\adsurl{http://cdsads.u-strasbg.fr/abs/2013ApJ...775L..23W}.
\end{barticle}
\endbibitem

\bibitem[\protect\citeauthoryear{{Wang} \textit{et~al.}}{2009}]{Wang2009}
\begin{barticle}
\bauthor{\bsnm{{Wang}}, \binits{T.J.}},
\bauthor{\bsnm{{Ofman}}, \binits{L.}},
\bauthor{\bsnm{{Davila}}, \binits{J.M.}},
\bauthor{\bsnm{{Mariska}}, \binits{J.T.}}:
\byear{2009},
\batitle{{Hinode/EIS observations of propagating low-frequency slow
  magnetoacoustic waves in fan-like coronal loops}}.
\bjtitle{\aap}
\bvolume{503},
\bfpage{L25}.
\doiurl{10.1051/0004-6361/200912534}.
\adsurl{http://cdsads.u-strasbg.fr/abs/2009A\%26A...503L..25W}.
\end{barticle}
\endbibitem

\bibitem[\protect\citeauthoryear{{Warren} \textit{et~al.}}{2011}]{Warren2011}
\begin{barticle}
\bauthor{\bsnm{{Warren}}, \binits{H.P.}},
\bauthor{\bsnm{{Ugarte-Urra}}, \binits{I.}},
\bauthor{\bsnm{{Young}}, \binits{P.R.}},
\bauthor{\bsnm{{Stenborg}}, \binits{G.}}:
\byear{2011},
\batitle{{The Temperature Dependence of Solar Active Region Outflows}}.
\bjtitle{Astrophys.J.}
\bvolume{727},
\bfpage{58}.
\doiurl{10.1088/0004-637X/727/1/58}.
\adsurl{2011ApJ...727...58W}.
\end{barticle}
\endbibitem

\bibitem[\protect\citeauthoryear{{Winebarger}
  \textit{et~al.}}{2002}]{Winebarger02}
\begin{barticle}
\bauthor{\bsnm{{Winebarger}}, \binits{A.R.}},
\bauthor{\bsnm{{Warren}}, \binits{H.}},
\bauthor{\bsnm{{van Ballegooijen}}, \binits{A.}},
\bauthor{\bsnm{{DeLuca}}, \binits{E.E.}},
\bauthor{\bsnm{{Golub}}, \binits{L.}}:
\byear{2002},
\batitle{{Steady Flows Detected in Extreme-Ultraviolet Loops}}.
\bjtitle{Astrophys.J.}
\bvolume{567},
\bfpage{L89}.
\doiurl{10.1086/339796}.
\adsurl{2002ApJ...567L..89W}.
\end{barticle}
\endbibitem

\end{thebibliography}

\end{article} 

\end{document}